\documentclass[apj,twocolumn,twocolappendix,dvipsnames]{aastex63}
\usepackage{epstopdf}
\usepackage{pbox}
\usepackage{gensymb}
\usepackage{amsmath}
\usepackage{amssymb}

\newcommand{\figurepath}{.}

\newcommand{\cs}{c_{\rm s}}

\newbox\grsign \setbox\grsign=\hbox{$>$}
\newdimen\grdimen \grdimen=\ht\grsign
\newbox\laxbox \newbox\gaxbox
\setbox\gaxbox=\hbox{\raise.5ex\hbox{$>$}\llap
     {\lower.5ex\hbox{$\sim$}}}\ht1=\grdimen\dp1=0pt
\setbox\laxbox=\hbox{\raise.5ex\hbox{$<$}\llap
     {\lower.5ex\hbox{$\sim$}}}\ht2=\grdimen\dp2=0pt

\shorttitle{3D jet launching}
\shortauthors{Sheikhnezami \& Fendt}

\begin{document}
\title{The physics of the MHD disk-jet transition in binary systems: \\  jetted spiral walls launched from disk spiral arms}
\author[0000-0002-4144-7373]{Somayeh Sheikhnezami}
\affiliation{Department of Physics, Institute for Advanced Studies in Basic Sciences (IASBS), P.O. Box 11365-9161, Zanjan, Iran}
\affiliation{School of Astronomy, Institute for Research in Fundamental Sciences (IPM), Tehran, 19395-5746, Iran}
\affiliation{Max Planck Institute for Astronomy, Heidelberg, Germany}
\author[0000-0002-3528-7625]{Christian Fendt}
\affiliation{Max Planck Institute for Astronomy, Heidelberg, Germany}
\email{snezami@iasbs.ac.ir, fendt@mpia.de}


     \date{\today}
\begin{abstract}
We present a detailed physical analysis of the jet launching mechanism of a circumstellar disk that is located in a binary system.
Applying 3D resistive MHD simulations, we investigate the local and global properties of the system, such as the angular momentum transport and the accretion and ejection mass fluxes.
In comparison to previous works, for the first time, we have considered the full magnetic torque, the presence of an outflow, thus the angular momentum transport by vertical motion, and the binary torque. We discuss its specific 3D structure, and how it is affected by tidal effects.
We find that the spiral structure evolving in the disk is {\em launched into the outflow}. 
We propose to call this newly discovered structure a {\em jet spiral wall}.
These spiral features follow the same time evolution, with the jet spiral somewhat lagging the disk spiral.
We find that the vertical transport of angular momentum has a significant role in the total angular momentum budget also in a binary system.
The same holds for the magnetic torque, however, the contribution from the $\phi$-derivative of magnetic pressure and the $B_{\phi}B_r$ stresses are small. 
The gravity torque arising from the time-dependent 3D Roche potential becomes essential, as it constitutes the fundamental cause for all 3D effects appearing in our disk-jet system.
Quantitatively, we find that the disk accretion rate in a binary system increases by $20\%$ compared to a disk around a single star. The disk wind mass flux increases by even 50\%.
\end{abstract}
\keywords{
   accretion, accretion disks --
   MHD -- 
   ISM: jets and outflows --
   stars: pre-main sequence, binary star --
   galaxies: jets --
   galaxies: active 
 }
\section{Introduction}
An exceptional phase during the early stellar evolution is the ejection of rapidly moving material into the interstellar medium (ISM) as collimated protostellar jets.
More general, jets are powerful signatures of astrophysical activity and are observed over a wide range 
of luminosity and spatial scale.
Besides young stars (YSO), also micro-quasars, and active galactic nuclei (AGN) are typical jet sources, while there is 
indication of jet motion also for a few pulsars and for gamma-ray bursts
\citep{1974MNRAS.167P..31F, 1979Natur.279..701A, 1983ApJ...274L..83M, 
1994Natur.371...46M, 1997ApJ...487L...1R}.

Astrophysical jets have been the subject of numerous studies investigating them from different points of view - such as the
process of jet launching, the jet propagation and interaction with the environment.
One of the earliest numerical simulations of radiatively cooling, supersonic jets was performed by \citet{1990ApJ...360..370B}.
Afterwards, the study of jet propagation using numerical simulations became feasible applying (M)HD codes developed by 
many groups \citep{1993ApJ...410..686D, 1993ApJ...413..198S,1993ApJ...413..210S,1994ApJ...420..237S}.
Regarding the jet feedback into the ambient gas, one of the first numerical simulation studying the impact of stellar outflows
on driving the interstellar turbulence was performed by \cite{2000ESASP.445..457M}.
Further studies on scales beyond the jet launching area and considering the interaction of the jet and the ambient gas 
were publishes subsequently (see e.g. \citealt{2007ApJ...668.1028B, 2009ApJ...692..816C, 2010MNRAS.402....7M, 2013MNRAS.429.2482P, 2014MNRAS.439.2903C, 2019ApJ...883..160S}.

On smaller scales, namely the jet formation and collimation scale, a break-through came by the simulations of \citet{1985PASJ...37...31S,1995ApJ...439L..39U,1997ApJ...482..712O}, numerically following the earlier, seminal 
analytical approaches by \citet{1982MNRAS.199..883B,1983ApJ...274..677P,1985PASJ...37..515U,Uchida1985}.
Such simulations considered the jet formation from the {\em disk surface}, thus the acceleration
of jet material and its collimation by the magnetic field 
(to cite a few, see \citealt{1993ApJ...410..218W, 1995ApJ...444..848L, 1997A&A...319..340F, 2002A&A...395.1045F,
2010ApJ...709.1100P, 2011ApJ...742...56V}) 

However, in order to understand the very launching process of the jet -- that is the transition from accretion 
to ejection -- it is essential to include the disk physics in the numerical treatment.
Today, numerical simulations of the accretion-ejection process play an essential role for the
understanding of jet launching.
A vast literature exists on magnetohydrodynamics (MHD) simulation on jet launching with ever improving physical complexity and also numerical resolution
\citep{Uchida1985, 1998ApJ...508..186K, 2002ApJ...581..988C, 2007A&A...469..811Z,2010A&A...512A..82M, 
2012ApJ...757...65S, 2014ApJ...793...31S, 2018ApJ...861...11S}.
In general, these works study how the properties of the outflow that is formed from the disk is determined from
certain disk properties, namely the disk resistivity, the presence of the mean field dynamo in the disk,
or 3D circum-stellar disk in a Roche potential.

Furthermore, we know that stars may form as binaries (see section below). 
In close binary pairs the axial symmetry of the jet source may be disturbed substantially.
Bipolar jets forming in a binary system may be affected substantially by tidal forces and torques,
that might be visible as 3D effects in the jet structure and jet propagation.

There are well-known observational signatures that strongly indicate on non-axisymmetric features like jet precession or a curved ballistic motion of the jet which are suggesting that the jet source is part of a binary 
system or even a multiple system 
\citep{Fendt1998, 2000ApJ...535..833S, 2002MNRAS.335.1100C, 2004HEAD....8.2903M, 2007A&A...476L..17A, 2014xru..confE.147M,
2016A&A...593A.132P, Beltran2016, 2019ASSP...55...71M,2019A&A...622L...3E, 2019IAUS..346...34M, 2021MNRAS.503..704M, 2021MNRAS.503.3145B, 2021MNRAS.tmp..799D}
These papers study different binary systems owing jets either from observational data or by applying 
simulation techniques.
Some of these jets are indeed found to show a non-axisymmetric structure, usually referred to as C-shape and S-shape,
which is thought to be a signature of jet precession of orbital motion of the jet source. 
All these features indicate the presence of binary or multiple system.

The launching of jets in a binary system and their subsequent propagation naturally requires a three-dimensional (3D) 
setup for the simulation.
The major difficulties here are
(i) the demand on CPU power, 
(ii) the different kind of physics for outflow and disk (ideal MHD or diffusive MHD, respectively), and 
(iii) the different time scales involved for the disk, the jet, and for the binary orbital motion.

Only recently, this could be achieved by \citet{2015ApJ...814..113S, 2018ApJ...861...11S} who tackled the problem of jet 
launching -- thus the accretion-ejection connection -- in ``3D simulations''.
The emphasis of these papers was on global properties such as the accretion and ejection mass fluxes, the overall 3D structure and stability of disk and jet, and on global tidal effects on disk and jet.
Here we continue these investigations, now concentrating on a much deeper consideration of the local and global effects of the angular momentum budget in the disk and the outflow.
In particular, we will investigate the effect of the existence of disk spiral arms for the launching process and the substructures emerging in the jets.

We will further investigate how the global observable such as disk accretion rate and jet outflow rate are affected in comparison to a single-star accretion disk that launches an outflow.
Similar works have been published, studying hydrodynamic torques in circum-binary disks 
\citep{1977MNRAS.181..441P, 1979MNRAS.186..799L, 2017MNRAS.468.1387L, 2017MNRAS.466.1170M,2019ApJ...875...66M, 2020A&A...635A.204A, 2020A&A...641A..64H}, 
or the torques exerted on accreting supermassive black hole binaries \citep{2013MNRAS.435.2633N} or the magnetic torque
in accretion disks of
millisecond pulsars \citep{2017MNRAS.469.4258T}.

Compared to previous studies of torques acting in a circum-binary disk of a binary system
(mostly performed in hydrodynamic limit),
our simulations consider the full magnetic torque and the presence of the MHD disk wind in a circum-primary disk in a binary system.
In general, we apply an approach similar to \citet{2017MNRAS.466.1170M, 2019ApJ...875...66M},
 meaning that we treat a similar set of equations, but apply them i) to a circum-primary disk, and 
extend them ii) in order to investigate the magnetohydrodynamic torques.  
With that, we can treat the launching and evolution of a disk jet from a circum-primary disk in a binary system.

 Note that in this paper, we concentrate on the evolution of a disk magnetic field and ignore the magnetic field of the 
central object that is subject to simulations of the disk-star interaction \citep{2013A&A...550A..99Z, 2018ApJ...857....4T,2019ApJ...878L..10T}.

Our paper is structured as follows.
In Section 2 we discuss the model setup for our simulations. 
We do this in brief, mostly referring to our previous papers \citep{2015ApJ...814..113S,2018ApJ...861...11S} in which
the modeling and numerical details are extensively discussed.
We then describe the general disk and outflow dynamics in Section 3 in great detail, discussing the particularities of the 3D disk and outflow dynamics.
Section 4 presents an analysis of the local torques acting in the disk and the outflow, 
while Section 5 discusses the global angular momentum budget.
In Section 6 we summarize the 3D effects concerning the mass and angular momentum fluxes.
Section 7 summarizes our paper.

For convenience we have compiled a table containing the various physical terms of the angular momentum budget that 
are considered and put it in Appendix A.
Additional useful information and graphs are included in Appendix B  and C.
\section{Model setup and equations}
This paper is the follow up work of our recent paper \citep{2018ApJ...861...11S} in which we consider a binary system with a {"}primary{"} of mass $M_{\rm p}$ and a {"}secondary{"} of mass $M_{\rm s}$, separated by the distance $D$.
The primary is surrounded by a disk of initial size $R_{\rm out} < D/2$.
The location of the secondary is chosen to be outside the computational domain.
The orbital plane of the binary system can be chosen to be inclined towards the initial accretion disk
by an angle $\delta$, however, for simplicity we do not consider this option for the present paper.
The Lagrange points L1, L2 and L3 are outside the initial disk radius.
The Lagrange points L1 and L3 could be located in the computational domain.
\subsection{Governing equations}
In the current paper we analysis the results of our 3D MHD simulations focusing on the physical process of jet launching
in the binary system. 
In these simulations we had applied the MHD code PLUTO \citep{2007ApJS..170..228M, 2012ApJS..198....7M}, version 4.3,
to solve the time-dependent, resistive, in-viscous MHD equations, 
accounting namely for the conservation of mass, momentum, and energy,
\begin{equation}
\frac{\partial\rho}{\partial t} + \nabla \cdot \left( \rho \vec u \right)=0,
\label{continuity}
\end{equation}
\begin{equation}
\frac{\partial \left( \rho \vec u \right) } {\partial t} + 
\nabla \cdot \left(  \rho \vec u \vec u \right) + \nabla P-\frac{ \left( \nabla \times \vec B \right) \times \vec B}{4 \pi}
+ \rho \nabla \Phi = 0.
\label{momentum_eq}
\end{equation}
\begin{multline}
 \frac{\partial e}{\partial t} + \nabla \cdot \left[ \left( e + P + \frac{B^2}{8\pi} \right) \vec u - \left( \vec u \cdot \vec B \right) \frac{\vec B}{4\pi} + \left( {\eta} \vec j \right) \times \frac{\vec B}{4\pi} \right]\\
 = - \Lambda_{\rm cool}. 
\end{multline}
Here, $\rho$ is the mass density, $\vec u$ is the velocity, $P$ is the thermal gas pressure,
$\vec B$ stands for the magnetic field, and $\Phi$ denotes the gravitational potential.
The electric current density $\vec j$ is given by Amp\'ere's law 
$\vec j = \left( \nabla \times \vec B \right) / 4\pi$.

The total energy density is
\begin{equation}
e = \frac{P}{\gamma - 1} + \frac{\rho u^2}{2} + \frac{B^2}{2} + \rho \Phi.
\end{equation}
We consider an ideal gas with a polytropic equation of state $P = (\gamma - 1) u$ with 
$\gamma = 5/3$ and the internal energy density $u$.
This is a further difference to \citet{2017MNRAS.466.1170M} and \citet{2019ApJ...875...66M}, both considering a locally 
isothermal gas.
This is a typical assumption for disk simulations, 
For studies of wind or jet launching simulations the literature usually assumes a polytropic.

The gas temperature is implicitly given by the polytrope, $T \propto \rho^{\gamma-1} \propto P/\rho$,
and is thus not considered as a separate variable.
Within our approach, heating (e.g. ohmic, compressional, numerical) will affect the dynamics via the gas pressure.

We consider a time dependent gravitational (Roche) potential $\Phi= \Phi_{\rm eff}$ in the equations.
Since the origin of our coordinate system is in the primary, we have to consider the time variation of 
the gravitational potential in that coordinate system.
We prescribe the position of the secondary initially $(t=0)$ along the $x$-axis.
Thus, its position vector varies over time as
\begin{equation}
\vec{D}= \hat{x} D \cos{\omega t} + 
         \hat{y} D \sin{\omega t}\cos{\delta} + 
         \hat{z} D \sin{\omega t}\sin{\delta},
         \label{roche_potential}
\end{equation}

with the inclination angle $\delta$ of the binary orbit with respect to the circum-primary disk.
Here $\hat{x}$, $\hat{y}$ and $\hat{z}$ denote the unit vectors in Cartesian coordinates.
In this paper we discuss a co-planar geometry, $\delta = 0$.
The effective potential in a binary system at a point with position vector $\vec{r}( x, y, z )$ is
\begin{equation}
\Phi_{\rm eff} = - \frac{G M_{\rm p}}{|\vec r|} - \frac{G M_{\rm s}}{|\vec{r}-\vec{D}|} 
                 + \frac{G M_{\rm s}}{|\vec{D}|^3}  \left(\vec{r} \cdot \vec{D}\right).
\label{eq:phi_eff}
\end{equation}
The first term in Equation~\ref{eq:phi_eff} is the gravitational potential of the primary,
while the remaining terms describe the tidal perturbations due to the orbiting secondary.
The last  {''}indirect{''} term accounts for the acceleration of the origin of the coordinate 
system (see also \citealt{1996MNRAS.282..597L,2018ApJ...861...11S}).

The evolution of the magnetic field is described by the induction equation,
\begin{equation}
\frac{\partial \vec B}{\partial t} - \nabla\times \left( \vec u \times \vec B - \eta \vec j \right) = 0.
\end{equation}

The magnetic diffusivity can be defined most generally as a tensor $\bar{\bar{\eta}}$
(see our discussion in \citealt{2012ApJ...757...65S}).
Here, for simplicity we assume a scalar, isotropic magnetic diffusivity as a function of space
$\eta_{ij} \equiv \eta(r,z)$.

The cooling term $\Lambda$ in the energy equation can be expressed in terms of ohmic heating
$\Lambda = g \Gamma$, with $\Gamma = ({\eta} \vec j) \cdot \vec j$, and with $g$ measuring 
the fraction of the magnetic energy that is radiated away instead of being dissipated locally. 
For simplicity, again we adopt $g=1$, thus we neglect ohmic heating for the dynamical evolution of
the system.
\subsection{Numerical specifics}
For the numerical specifics such as boundary conditions, initial conditions, and the numerical grid we refer to our 
previous paper \citep{2018ApJ...861...11S}.
Here we want to emphasize that all simulations were performed applying Cartesian coordinates.
This is essential in order to exclude any artificial effect of the rotational axis on the 3D structure
of the outflow.

However, in the present paper we will mainly discuss the evolution of properties involving radial motions (accretion, ejection)
and toroidal motions (orbital motions, angular momentum and torques with respect to the original rotational axis).
In particular, we also need to integrate in $\phi$-direction.

We therefore need to transform the required physical variables from the Cartesian to a cylindrical coordinate system.
As it is well known this transformation may provide some traps arising from the treatment of the trigonometric functions in the four quadrants.
We have therefore thoroughly tested our transformation routines in order to deal actually with the proper physical quantities.

We have also applied the interpolation tool provided by PLUTO to interpolate the variables that were 
evolved by the simulation on a Cartesian grid onto a cylindrical coordinate system.
This option was used in particular when we further needed to integrate global properties such as the disk angular
momentum at a certain radius, or when plotting variables along the azimuthal angle $\phi$.
\begin{figure*}
\centering
\includegraphics[width=18cm]{\figurepath/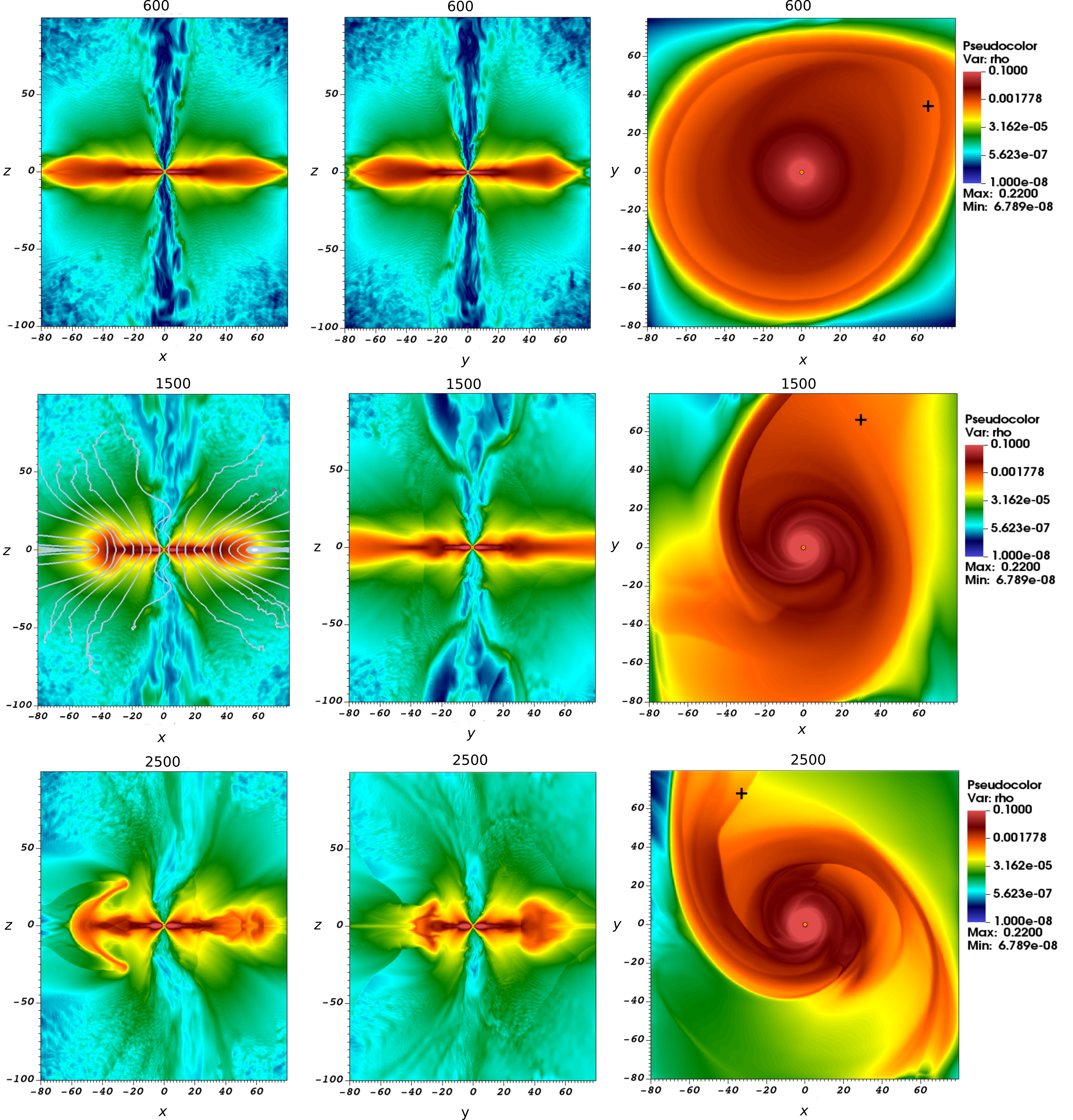}
\caption{Time evolution of the disk-jet structure in a binary system.
Shown are snapshots of the mass density in log scale for simulation run {\em a0} at $t = 600, 1500, 2500$ in the $xz$-plane, 
in the $yz$-plane, and in the mid-plane $z=0$.
The position of the $L_1$ point is indicated with the {``}+{''} sign.}
\label{fig:nc3_xy_rho_com}
\end{figure*}
\subsection{Units and normalization}
The simulations are performed in code units.
To convert in astrophysical units, for a protostellar system we may apply for the inner disk radius 
$r_{\rm i} = {0.1~\rm au}$ and the Keplerian velocity at the inner disk radius $u_{\rm K,i}$,
result in a dynamical time scale of 
\begin{equation}
t_{\rm i} = \frac{r_{\rm i}}{u_{\rm K,i}}
          =  1.8 \left( \frac{r_{\rm i}}{0.1\rm au} \right)^{3/2}
             \left( \frac{M_{\rm p}}{M_\odot} \right)^{-1/2} {\rm days }.
             \label{eq:time-unit}
            \end{equation}
Thus, a running time of the simulation of $5000\,t_{\rm i}$ corresponds to 25 years for a 
typical protostellar system.

Other systems of interacting binaries are Cataclysmic Variables (CVs), consisting of a white dwarf (WD) 
as a primary and a late type main sequence star as secondary in close separation.
The secondary may serve as source of material that is accreted onto the primary via an accretion disk.
The typical orbital period observed for CVs is about a few hours.
In order to scale our simulations to a CV system we may choose an inner disk radius to be several WD radii, 
thus $r_{\rm i} \simeq 5\times 10^4 {\rm km}$ \citep{2016AstL...42..379S}.

The astrophysical time scale of our simulations applied to CVs is 
\begin{equation}
 t_{\rm i} = 0.85 \left( \frac{r_{\rm i}}{5\times10^4\rm km } \right)^{3/2}
                    \left( \frac{M_{\rm p}}{M_\odot} \right)^{-1/2} {\rm hours. }
\end{equation}

 More details on the normalization of the variables are provided in Appendix C.

\begin{table} 
\caption{Characteristic simulation parameters: 
initial (maximum) plasma-beta at the inner disk radius,$\beta_{\rm i}$,
binary separation $D$, 
inclination angle between binary orbit and the disk mid-plane,$\delta$,
mass ratio between secondary and primary, $q\equiv M_{\rm s}/M_{\rm p}$,
radial location of the Lagrange points (orbital plane) $r_{\rm L1}$, $r_{\rm L3}$, 
and the orbital period $T_{\rm b}$  (in units $t_{\rm i}$).
 The initial aspect ratio of the disk is $\epsilon = 0.1$,
and the initial outer disk radius $r_{\rm out} = 65$.
All values are given in code units.}
\begin{center}
\begin{tabular}{lccccccccl}
\hline
\hline
\noalign{\smallskip}
Run  & $\beta_{\rm i}$ & D & $\delta$ & $ q $ & $r_{\rm L1}$ & $r_{\rm L3}$ & $T_{\rm b}$ \\
\noalign{\smallskip}
\hline
\noalign{\smallskip}
\noalign{\smallskip}
{\em a0}        & 20  & 150          &  0   & 1   &  75 &  105 & { 8160}  \\
{\em a1 }       & 20  & single  & -    & -   &  -  &  -   &   -      \\
  \noalign{\smallskip}
 \hline
 \noalign{\smallskip}
 \end{tabular}
 \end{center}
\label{tbl:0}
\end{table}
\section{A 3D jet launching disk simulation}
In this section we present and discuss the different physical variables of 3D MHD simulations of jet launching in binary systems.
For details we refer to our past publications \citep{2015ApJ...814..113S,2018ApJ...861...11S}.

We first briefly summarize the general evolution of the accretion-ejection structure by discussing our reference simulation
{\em a0} for which the binary orbit and the disk forming jet are co-planar
(see Table \ref{tbl:0}).

This simulation shows all the tidal effects caused by the secondary, but not the effect of a disk re-alignment and a subsequent
3D disk or jet precession, as this requires an inclination between jet-launching disk and orbital plane..

As a general feature of the disk evolution, we find that the disk size is decreasing and becomes finally 
confined to a size within the Roche lobe.
From an axisymmetric initial state the disk evolves into an asymmetric structure after $t=500$, developing
a spiral arm structure that grows in time.
\begin{figure*}
\centering
\includegraphics[width=18cm]{\figurepath/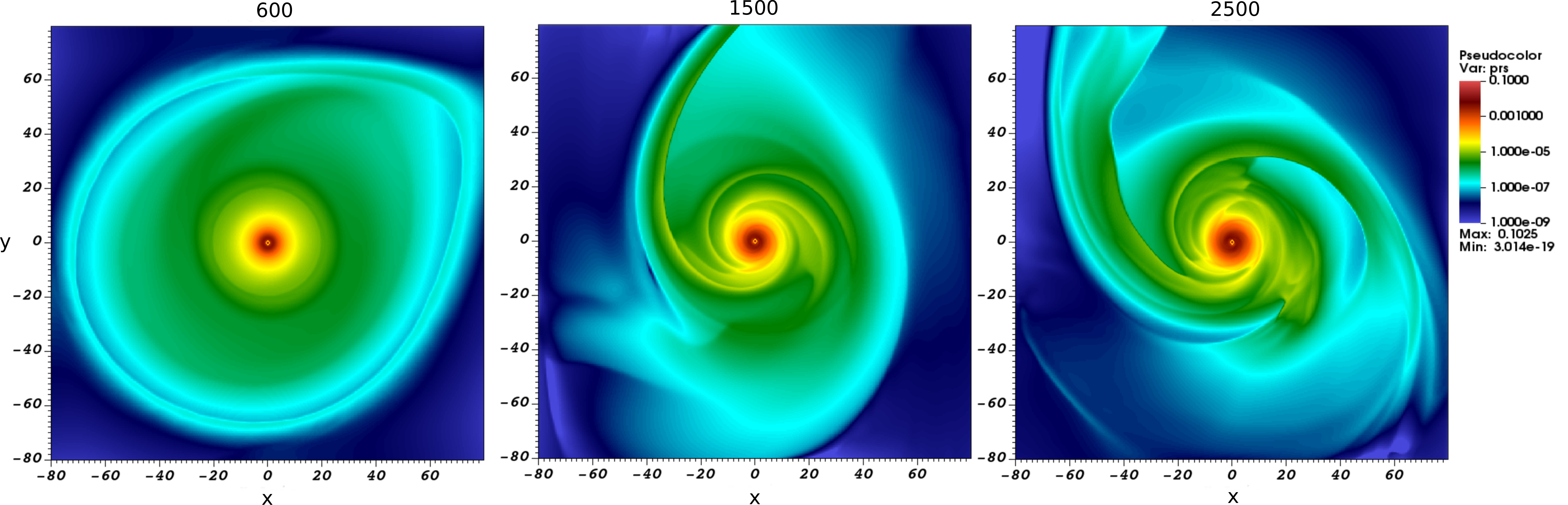}
\caption{Time evolution of the disk-jet structure in a binary system.
Shown are snapshots of the gas pressure in log scale for simulation run {\em a0} at $t = 600, 1000, 2500$ in the mid-plane $z=0$.
}
\label{fig:nc3_xy_prs_com}
\end{figure*}

\subsection{Evolution of the disk spiral arms}
We now consider in particular the evolution of the disk spiral arms for our reference run {\em a0}.
We first look at snapshots of the mass density (Fig.~\ref{fig:nc3_xy_rho_com}) and the gas pressure (Fig.~\ref{fig:nc3_xy_prs_com})
across the equatorial plane, $z=0$, for three exemplary time steps.
The spiral arm pattern appears at the same position in both maps, respectively -- simply indicating that 
the density wave and the pressure wave follow the same pattern speed.

We notice that the spiral arms first evolves smoothly as a density wave but then develops 
a shock structure with a jump in pressure and mass density.
The shock front allows for a clear definition of the spiral arm position.
The gas accumulates at the shock front making the spiral arm structure more prominent over time. 
It is clearly seen that with time the disk spiral arms become denser and more prominent and represent
the main structural feature of the disk.
The spiral arms rotation is synchronized with the orbital motion of the binary.

The magnetic field lines of the accretion-ejection system are shown in Figure~\ref{fig:nc3_xy_rho_com}
for $t=1500$ for the $xz$-plane.
It shows a smooth, almost axisymmetric pattern that is typical for time scales up to $t=1500$. 
However, when the disk spiral arms become more prominent,
we will see the 3D effects of the dynamical evolution more clearly
(for reference, see \citealt{2018ApJ...861...11S}).

In order to analyze the motion and the pattern speed of the spiral arms and the gas material during the evolution of the accretion disk,
we first follow the evolution of the density peaks inside the spiral arms.
The  ({"}northern{"}) spiral arm starts forming at $t \simeq 1000$, while at the opposite side, the signature of 
a spiral arm appears somewhat later, at $t \simeq 1500$.
We believe that this time difference arises from the fact that {"}southern{"} part of the disk is just farther to the companion star and tidal forces that form the spiral wave are weaker and thus need more time to evolve.
Also, the {"}northern{"} is directed towards the secondary (see the position of the L1 moving),
and thus feels the tidal forces stronger.

\subsection{Disk and outflow dynamics}
In order to gain insight in the evolution of the spiral arm pattern we now consider the dynamics of the accretion disk and the outflow in more details.

\begin{figure}
 \includegraphics[width=1.\columnwidth]{\figurepath/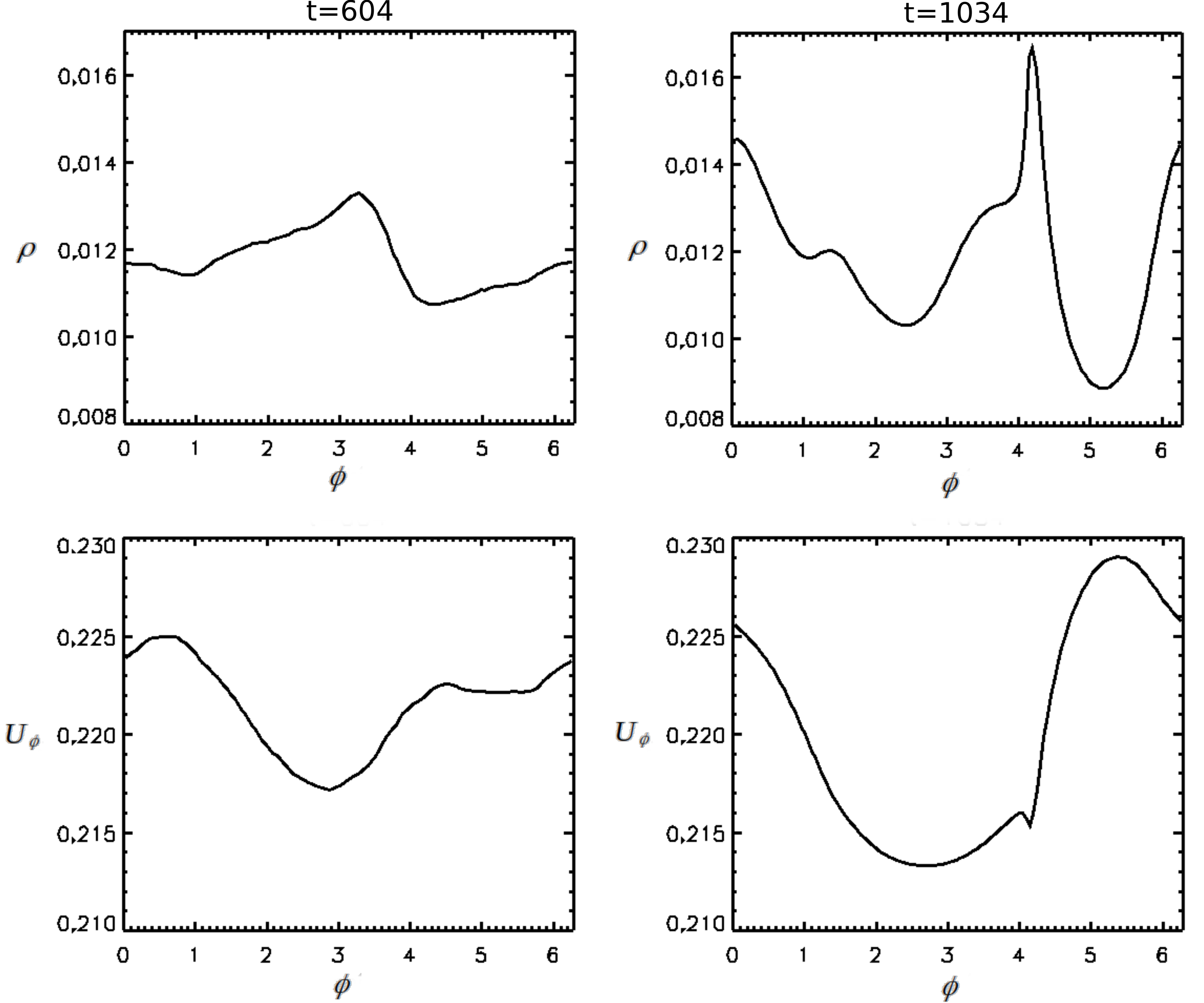}
\caption{Angular profiles of density and rotational velocity at $r=20$.
Shown are mid-plane values at times $t=604, 1030$.}
\label{fig:vpatt}
\end{figure}

We first consider the angular profile of mass density $\rho(\phi)$.
For our numerical estimates we consider Figure~\ref{fig:vpatt} (first row).
Similarly, Figure~\ref{fig:vpatt} (bottom row) shows also the angular profile of the rotational 
velocity $u_{\phi}$ for $r=20$.

The spiral arms are clearly detected by the peaks and dips in the corresponding angular profiles.
By comparing the location of these features over time we can estimate the pattern speed of the spiral arm.

When comparing the profile of mass density and the rotational velocity along circles of different radius, $\rho(\phi; r)$, and $u_{\phi}(\phi; r)$ (not shown), we clearly observe that these profiles differ for different radii.
Furthermore, we find that the density peak(s) at a certain radius move in angular direction over time, indicating the rotation 
of the spiral arm pattern. 
For instance, at $t= 604$ the mass density peaks at larger radii, $r>5$,
indicating that the arm is gradually forming at larger radii we do not show the $\phi$ profile for all radii). 
Evidently, this once again shows the spiral structure of the arm. 

We now derive some numerical estimates, considering Figure~\ref{fig:vpatt}.
From the motion of the density peaks we can derive the pattern speed of the spiral arm(s),
\begin{equation}
  u_{\rm patt} = r \frac{\Delta \phi}{\Delta t}.
\end{equation}

For $r= 20$ we estimate the pattern speed $u_{\rm patt}$ focusing on the mass density (mid-plane) at times 
$t=604$ and $t=1034$.
We find the peak of the density profile moving in $\phi$ direction from $\phi_1=3.3$ to $\phi_2=4.2$,
resulting in a circular pattern speed $u_{\rm patt} = 0.041$.

The Keplerian velocity at the disk mid-plane at $r=20$ is $u_{K} = 0.22$ and, thus one order of magnitude larger than the pattern speed at this radius.
This is typical for any orbiting wave pattern, while the exact pattern speed of course depends on the forcing involved \citep{1964ApJ...140..646L} .
Here the wave pattern is triggered by the orbiting secondary with an orbital period of about 10,000 time units.
Overall, this looks all reasonable and shows again that the arm is indeed a pattern that is not co-rotating with the 
material, but is synchronized with the companion motion.

\begin{figure*}
\centering
\includegraphics[width=18cm]{\figurepath/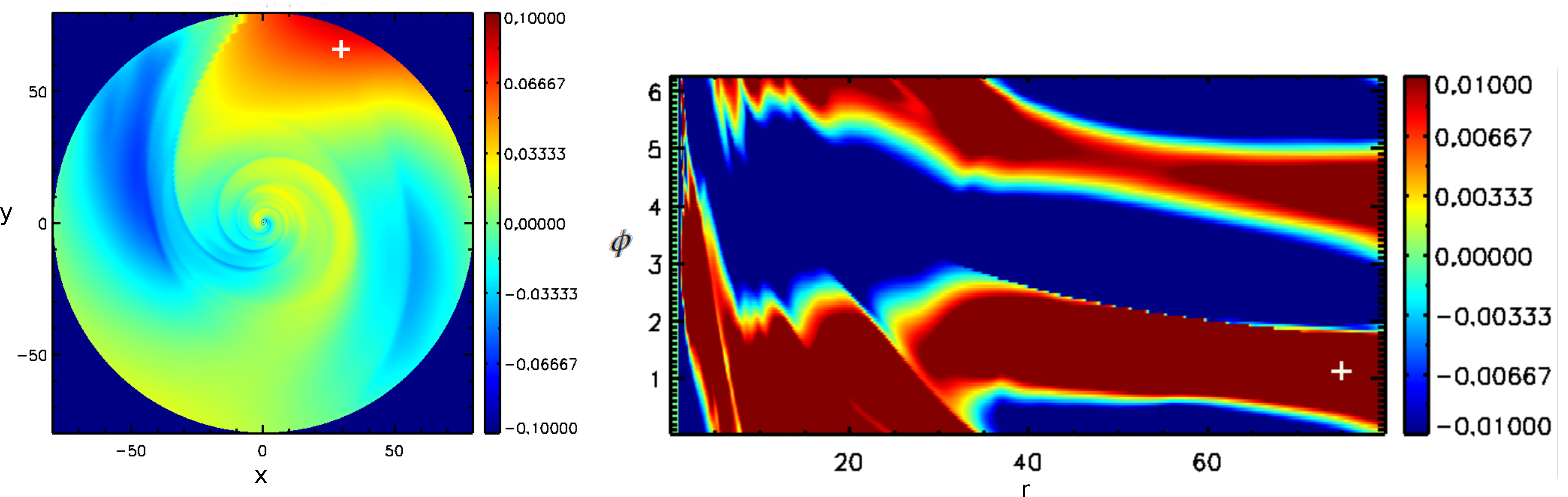}
\caption{Radial velocity field $u_r(r,\phi)$ in the disk mid-plane at $t=1500$.
For convenience, two representations are shown.
The standard image in Cartesian coordinates (left) emphasizes the spiral structure of the velocity field.
For the image that is shown in cylindrical coordinates (right) we have chosen a color bar that emphasizes the inflow-outflow
structure along the mid-plane.
The '+' symbol indicates the position of the L1 point.}
\label{fig:vr_binary2}
\end{figure*}

\begin{figure*}
\centering
\includegraphics[width=18cm]{\figurepath/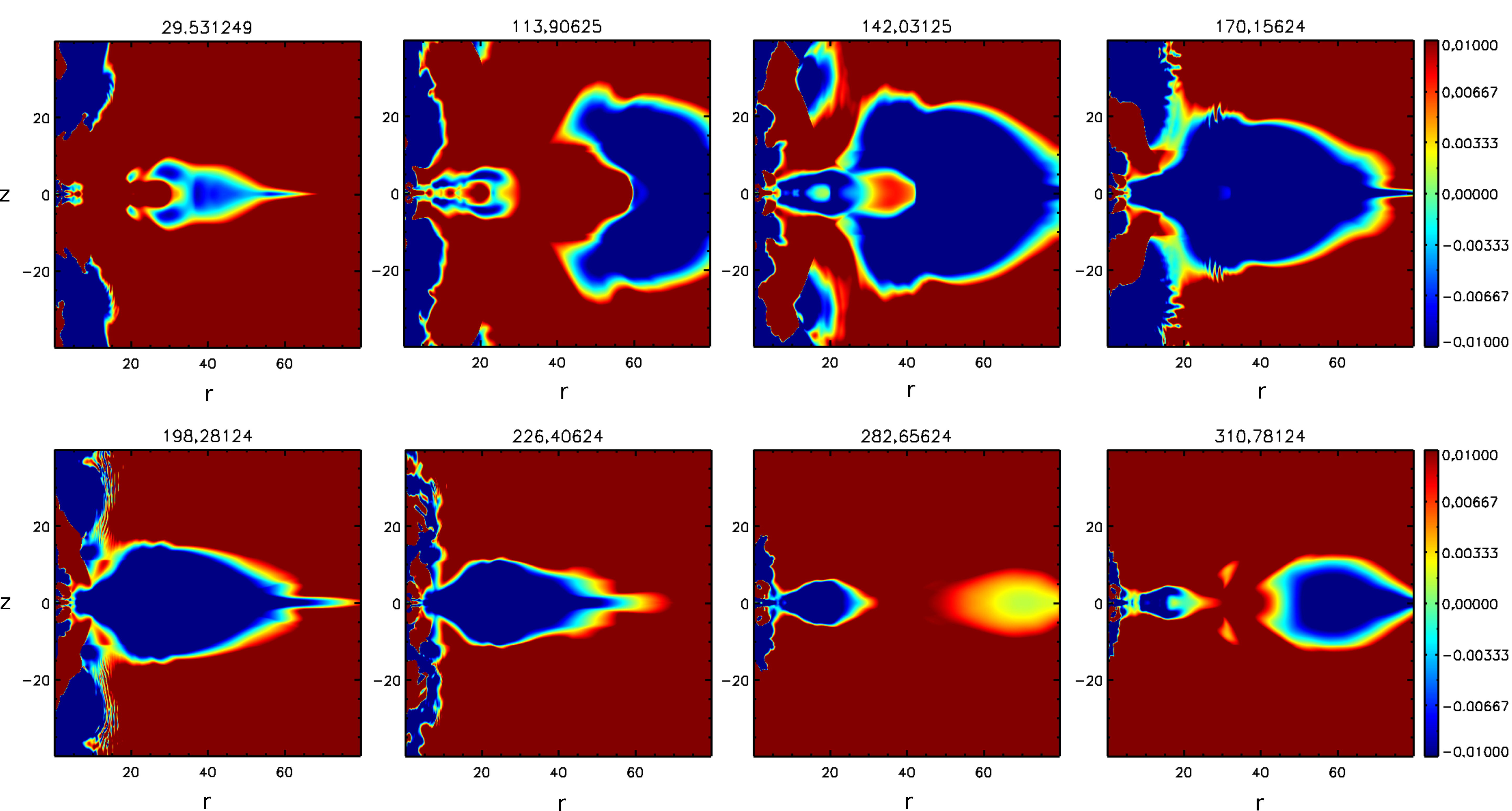}
\caption{Accretion-ejection velocity field. Shown are the snapshots of the radial velocity $u_r$.
Compare to Fig.~\ref{fig:nc3_xy_rho_com}, middle, for a face-on view of gas density. 
The angle $\Phi =0$ is measured from the $x$-axis. 
Different $r$-$z$ planes correspond to different angles $\phi$.
In Figure~\ref{fig:vr_single} we show for comparison the $u_r$-distribution for a single star simulation.
Colors are enhanced to demonstrate inflow (blue) and outflow (red).
}
\label{fig:vr_binary}
\end{figure*}

We can now quantify the growth rate of the spiral arms by simply comparing the height and width of the density peak over time or more accurately, by integrating the density profile over the width of the arm.
For this we look again at Fig.~\ref{fig:vpatt} and estimate how the peaks in the density profile grow over time.
As an example we consider the spiral arm located at $\phi = 3.3$ at $r=20$ at $t=604$, which is moving to $\phi =4.2$ at $t=1034$.
 
We integrate the mass under the density peaks within a control volume, here defined by the 
two minima along $\phi$, by $\Delta r =1$, and integrating from $z=-1$ to $z=1$, thus 
 \begin{equation}
 \Delta M = \int_{r=20}^{r=21} \int_{\phi_{\rm min1}=2.4}^{\phi_{\rm min2}=5.2} \int_{z=-1}^{z=1} \rho~r~d\phi~dz~dr.
  \label{secondinteg}
 \end{equation}
Since we look for an estimate only, we consider the high density area close to the mid-plane.
We refer to the growth rate as to the {\em local excess} mass that is carried by the spiral arm, meaning the total mass enclosed by the arm
(in the control volume with $\Delta r=1$),
but subtracted by the average disk mass.
Here, we consider the azimuthally averaged density as proxy for the underlying disk.

We thus measure a {\em local} growth rate of the excess mass of the arm of
$\dot{M}_{\rm{arm}} = {\Delta M}_{\rm{arm}} / \Delta t = 0.22 / 430 = 5 \times10^{-4}$ in normalized units,
corresponding to $\dot{M}_{\rm{arm}} =0.22/0.77= 0.285$ measured per Keplerian period at $r=20$. 

This numbers make sense only when compared to the local disk mass in the control volume, 
$\Delta M =2.95$, thus referring to a growth rate of 10\% over a Keplerian period at $r=20$.
We note that the timescale of the growth rate is consistent with the sound crossing time across the disk, $H/R \simeq 0.1$.

Naturally, the same numbers hold for the growth rate of the disk density.
Note, however that disk mass decreases (and thus the mean disk density) over long time due to 
ongoing accretion and ejection (see our discussion in \citealt{2018ApJ...861...11S}).

Now we discuss the overall pattern of the radial velocity field. 
Figure~\ref{fig:vr_binary2} shows the radial velocity distribution in the disk mid-plane, $u_r(r,\phi)$. For convenience, two representations are shown.
The standard image in Cartesian coordinates (left) emphasizes the spiral structure of the velocity field.
For the image that is shown in cylindrical coordinates (right) we have chosen a color bar that emphasizes the inflow-outflow 
structure along the mid-plane.

We see that there are separate streams of opposite radial direction in the disk mid-plane.
In particular, we recognize that close to the spiral arm the direction of the gas materials changes. Interestingly, we find a positive radial motion (red area, left panel) in direction of the secondary. This area is outside the Roche lobe (compare to the position of the L1 in Fig.~\ref{fig:nc3_xy_rho_com}).
Further, material is spiraling in (blueish color) along the {"}left{"} spiral arm, and moving
out along the {"}right{"} spiral arm (yellowish color).

In Figure~\ref{fig:vr_binary} we display the radial velocity field of the disk-jet structure in the meridional plane.
Although the radial velocity pattern looks quite unusual, this is not an artifact of our data handling. We may compare this to the 3D simulation results of a single star (see Appendix, Figure~\ref{fig:vr_single} which shows a regular accretion pattern (a negative $u_r$) in almost axisymmetry.
In contrary, the accretion velocity for the binary star simulation looks drastically different (see Figure~\ref{fig:vr_binary}). 
Here, positive radial velocities exist in the disk, indicating {"}excretion{"} channels along certain angular directions.
These channels are most clearly indicated in Figure~\ref{fig:vr_binary2} (right panel) that clearly shows radial layers of inverse radial velocity.
Accretion happens (at this time $t=1500$) along $\phi = 90\degr, 270\degr$, while excretion dominates in channels along $\phi = 0\degr, 180\degr$.
Most probably given by the spiral geometry of the disk structure, these channels are not completely aligned along the radial direction (thus not constant along certain angles).
Indeed these channels also follow a spiral structure, as in Figure~\ref{fig:vr_binary2} they are not oriented parallel to the horizontal axis. 

We now discuss the rotational velocity distribution.
Comparing the angular profiles of the rotational velocity $u_{\phi}(\phi)$, we recognize that 
these profiles change for different radii. 
In Figure~\ref{fig:vpatt} we show the angular profile of rotational velocity at $r=20$. 

While the disk material follows more or less a constant rotation profile along $\phi$ for small radii
$r=5, 15$, for larger radii this profile is substantially different (not shown $\phi$ profile for all radii).
The profile of rotational velocity follows very closely the profile of the density (see Fig.~\ref{fig:vpatt})
Peaks in the rotational velocity profile indicate the location of the spiral arm, while these peaks also indicate a very strong shear.

Enhanced {\em orbital} velocity that is present in the disk, itself triggers further angular momentum exchange 
(and also heating in case of a viscous approach that we do not follow).
We also observe a combination of super and sub-Keplerian velocities
that reflects different behaviors of mass flow in the disk.
Concerning the angular momentum balance, the material in super-Keplerian regions has gained angular momentum.
Quoting \citet{2001LNP...573...69B}, we stress that spiral waves in disks carry a negative angular momentum and their dissipation
leads to accretion of the fluid supporting the waves onto the central object.
 This issue has been addressed also before, see e.g.~\citet{2016ApJ...823...81J}. 

\subsection{Spiral arms injected into the outflow}
In the last section, we have analyzed the structure and evolution of the disk spiral arms.
We now consider how this structure that is generated in the disk, is further transferred into the outflow.

\begin{figure*}
\centering
\includegraphics[width=18cm]{\figurepath/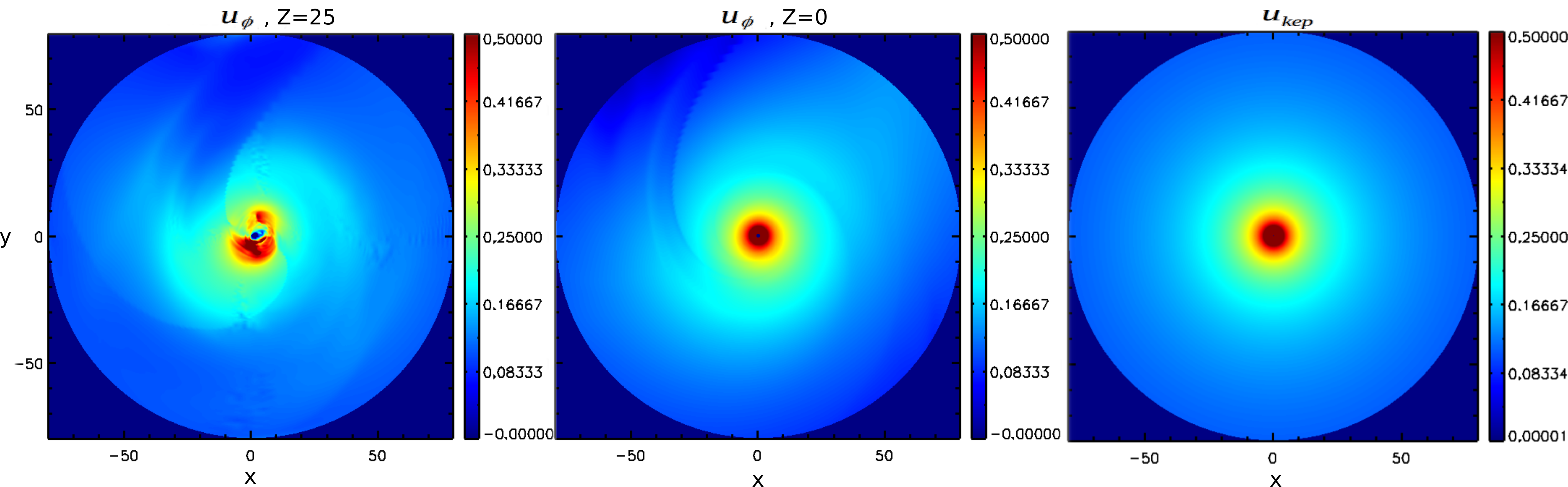}
\caption{Velocity field in the disk and the jet at $t=1500$.
Shown is the rotational velocity $u_{\phi}$ of the jet material (at $z=25$) and in the disk mid-plane.
For comparison we display also the Keplerian velocity $u_{\rm K}$ at the disk mid-plane at t=1500.}
\label{fig:vtor1500}
\includegraphics[width=18cm]{\figurepath/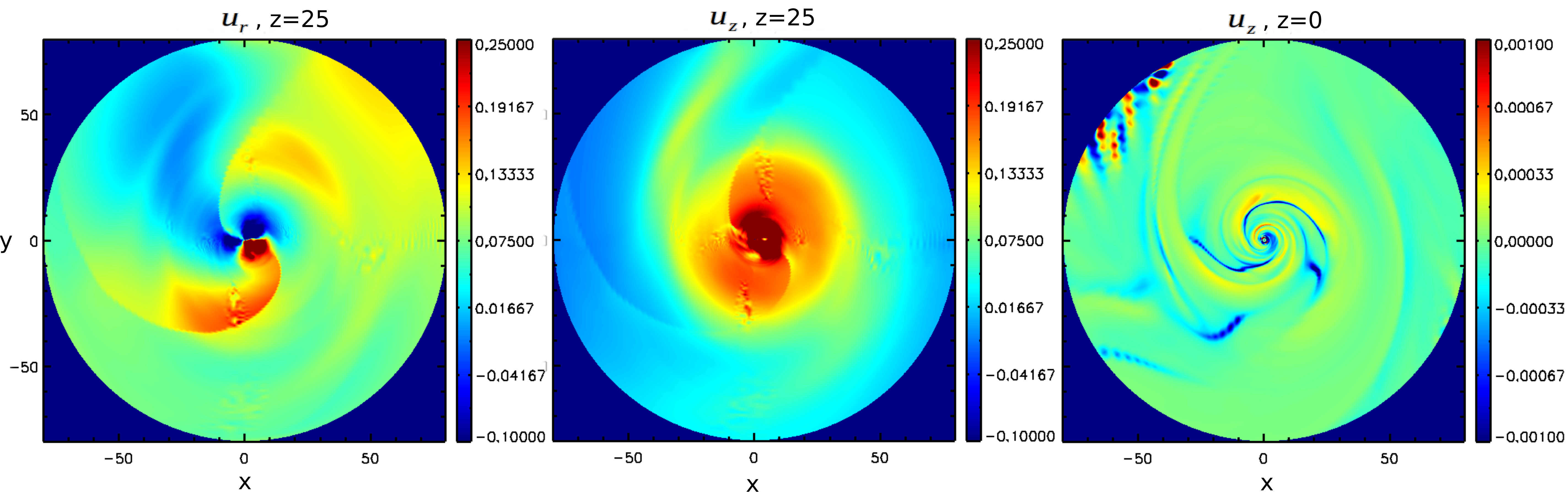}
\caption{Velocity field in the disk and the jet at $t=1500$.
Shown are the radial velocity $u_r$ of the jet material (at $z=25$) and the vertical velocity $u_z$ of 
the jet material and  at the disk mid-plane.}
\label{fig:vr1500}
\end{figure*}

 \begin{figure*}
 \centering
\includegraphics[width=18cm]{\figurepath/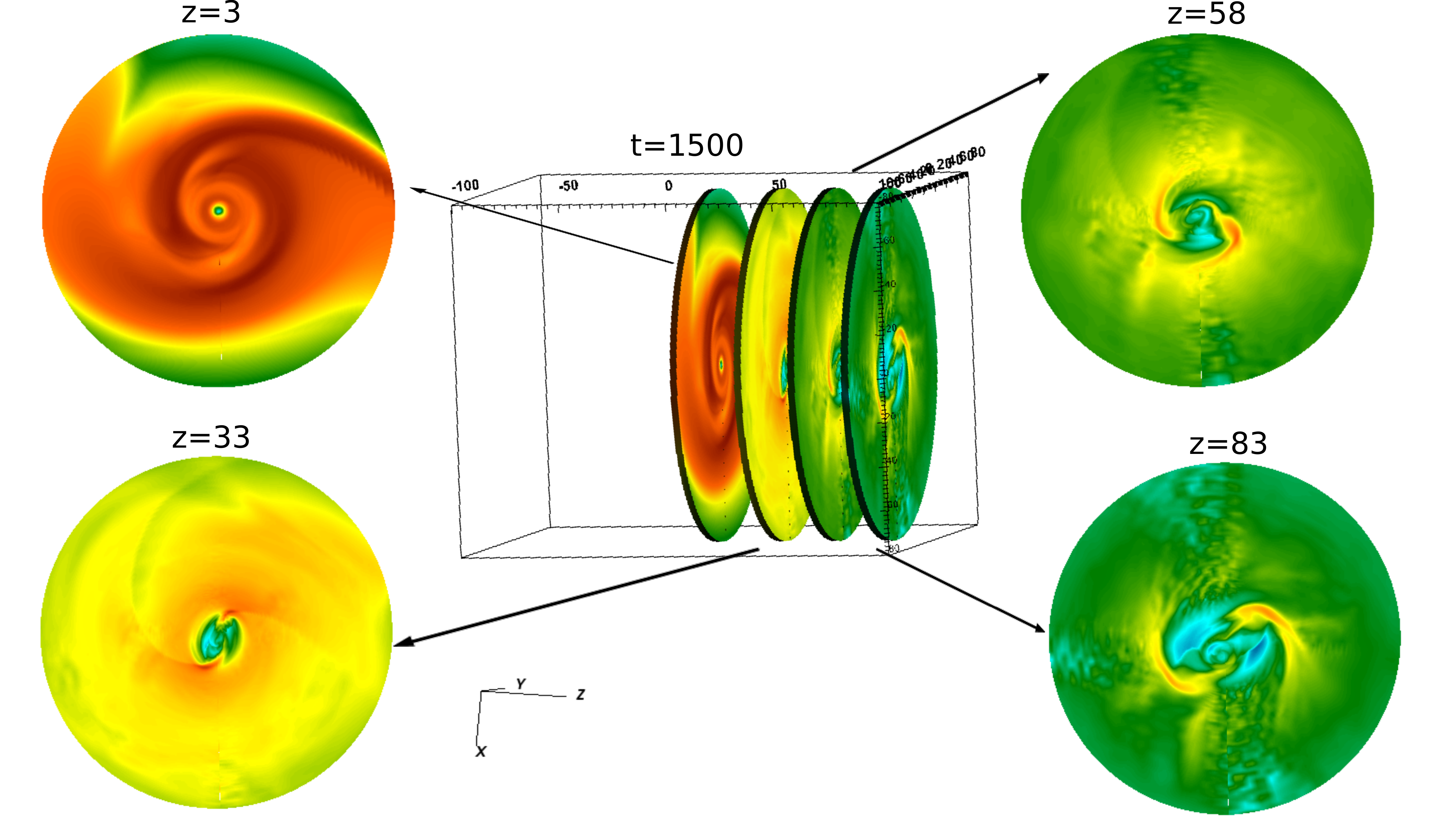} 
\caption{Evolution of the spiral structure injected into the jet. 
Shown are the 2D slices of a 3D snapshot at time $t=1500$ of the density (in log scale) 
at different height within the disk-jet system, $z=3$, $z=33$, $z=58$ and $z=83$.
At this time the spiral wall is fully developed with an angular shift of the spiral geometry 
between the different layers, corresponding to a time lag caused by the jet propagation.
}
\label{fig:3dview_box_diffz}
\end{figure*} 

\begin{figure}
 \includegraphics[width=1.0\columnwidth]{\figurepath/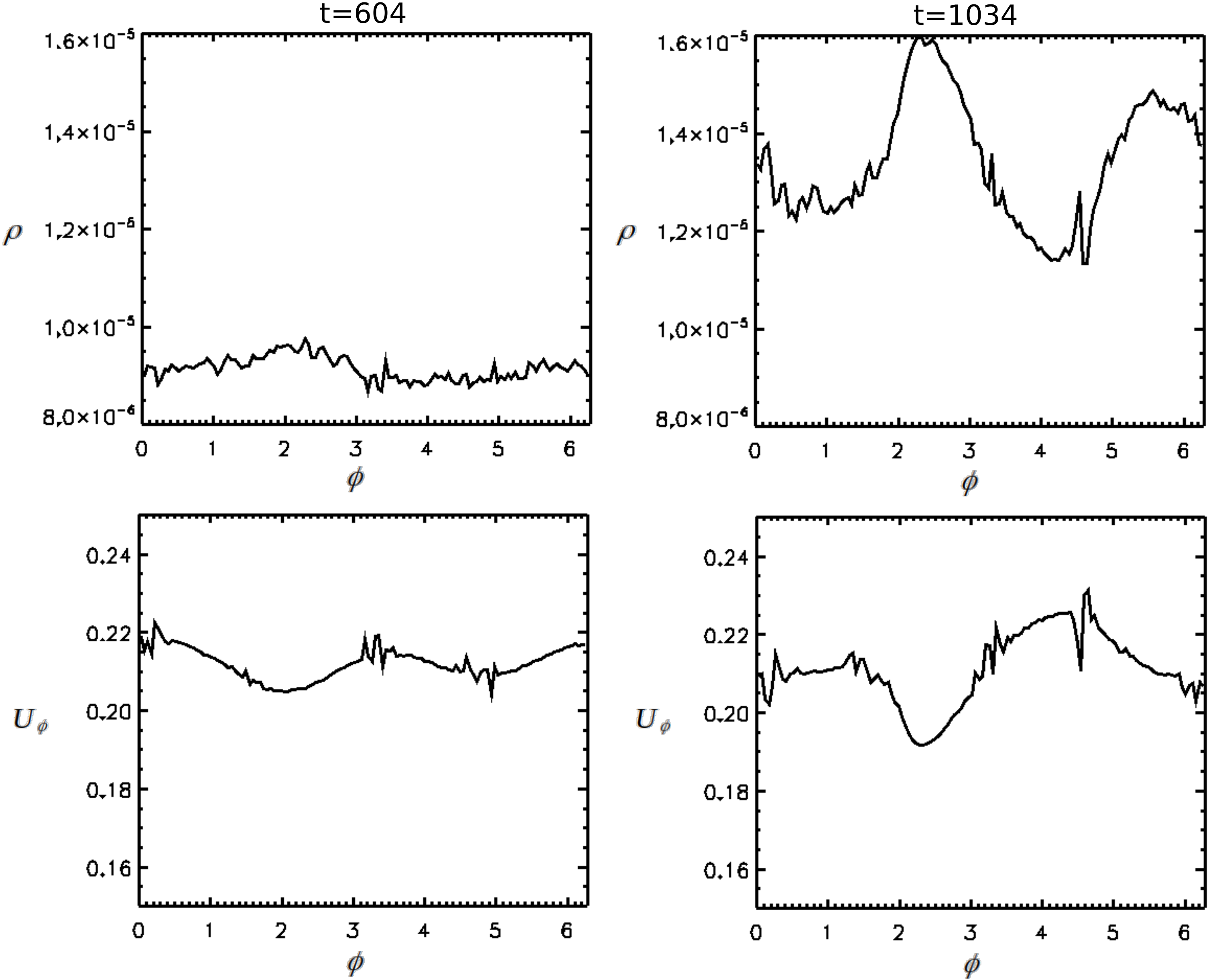}
\caption{Angular profiles of density and rotational velocity at $r=20$ in the jet (at $z=25$)
for two different evolutionary time steps, $t=600, 1030$. }
\label{fig:vpatt-jetz30}
\end{figure}

\begin{figure*}
 \includegraphics[width=18cm]{\figurepath/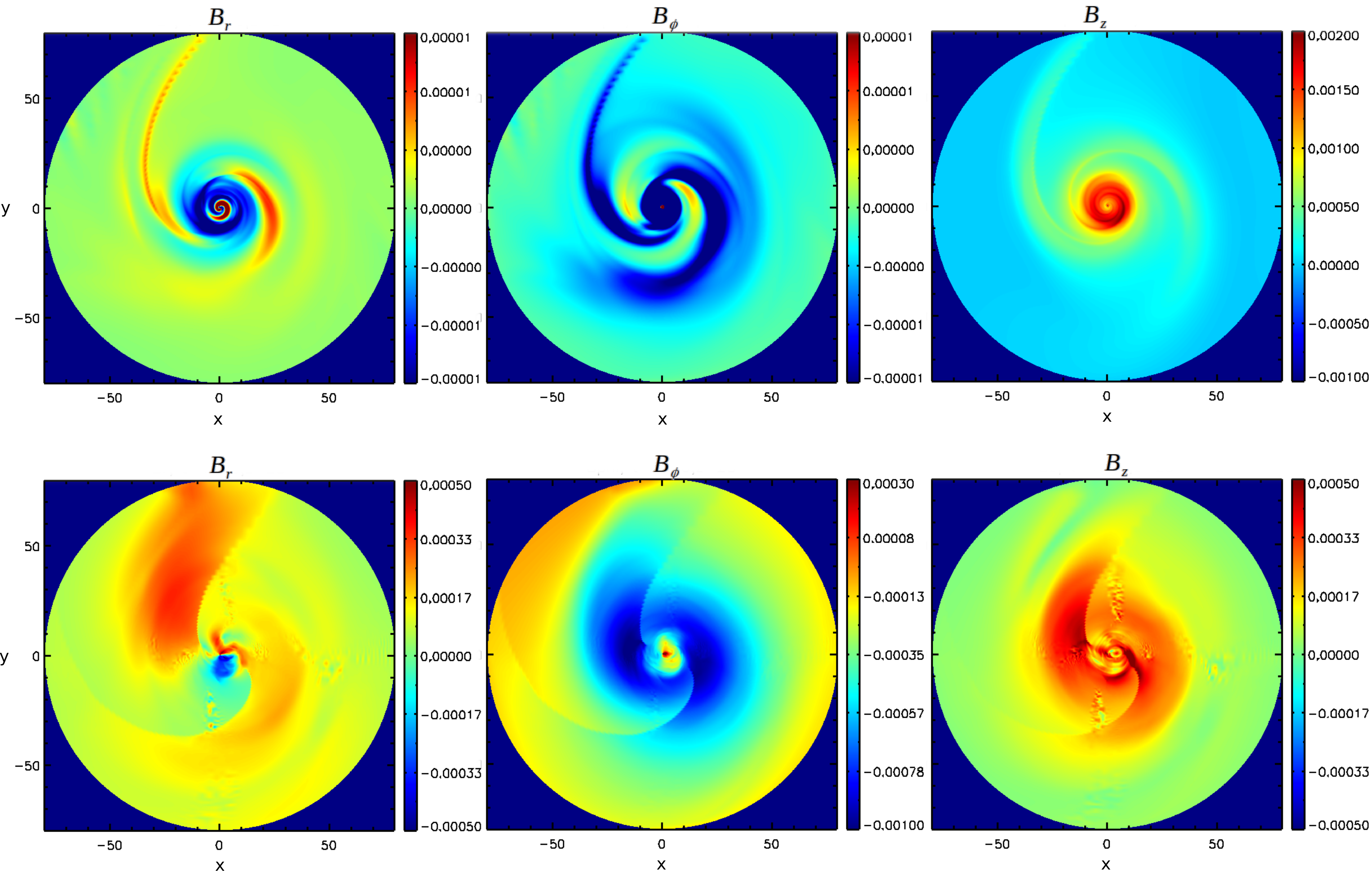}
\caption{Magnetic field components in the disk-jet structure at $t=1500$. 
Shown are the radial, the vertical and the toroidal component of the magnetic field  
in the disk mid-plane $z=0$ (top), and the jet material at $z=25$ (bottom). }
\label{fig:brbphibzjetdisk}
\end{figure*}

\begin{figure*}
\centering
\includegraphics[width=18cm]{\figurepath/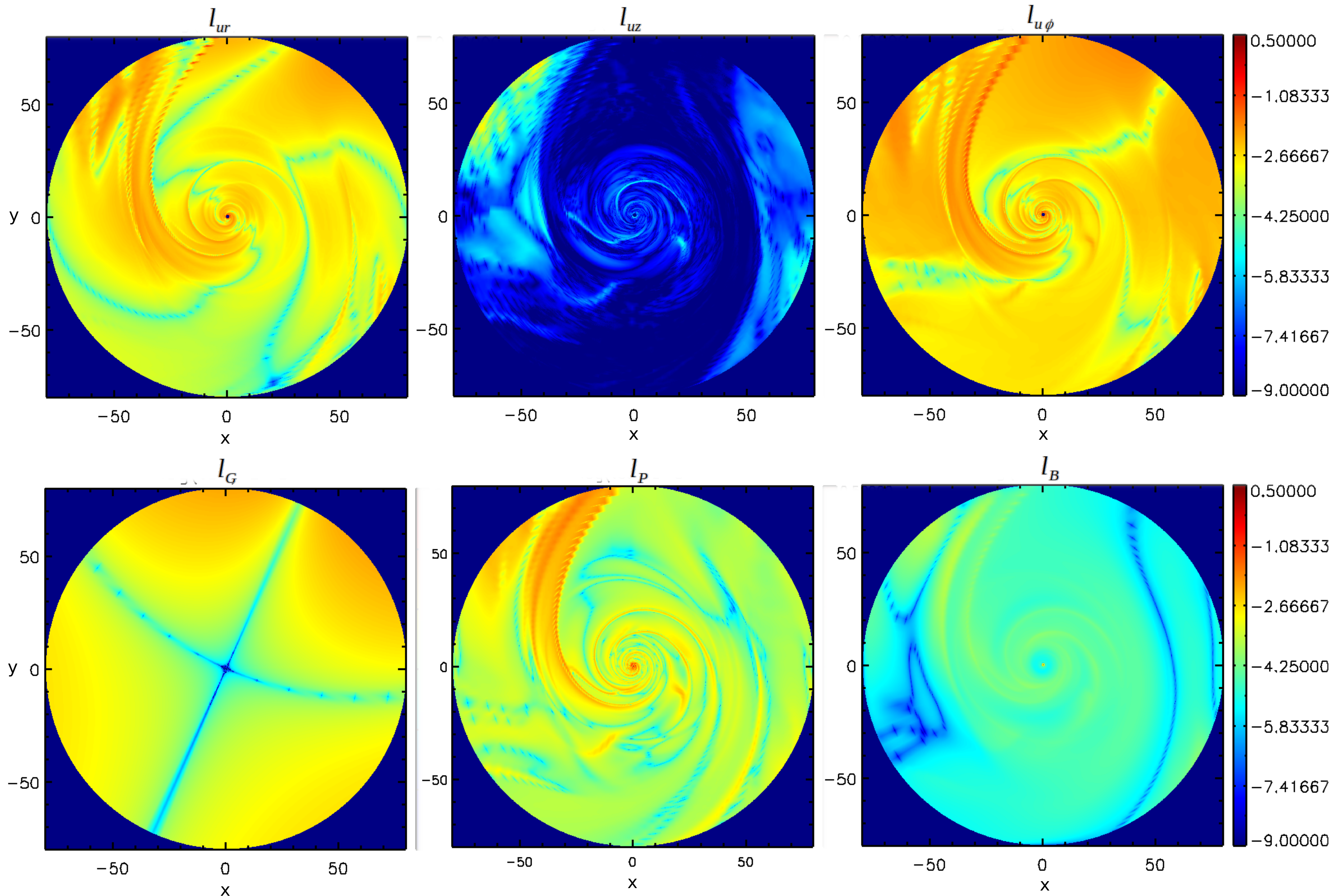}
\caption{Specific angular momentum distribution in the disk mid-plane at time $t=1500$ in log scale.
Shown are all terms contributing to equation~\ref{llintime} in consecutive order (from upper left to lower right).
For comparison these terms are displayed in
Table~\ref{tbl:terms} (top).
}
\label{fig:eq16_disk}
\end{figure*}

\begin{figure*}
\centering
\includegraphics[width=18cm]{\figurepath/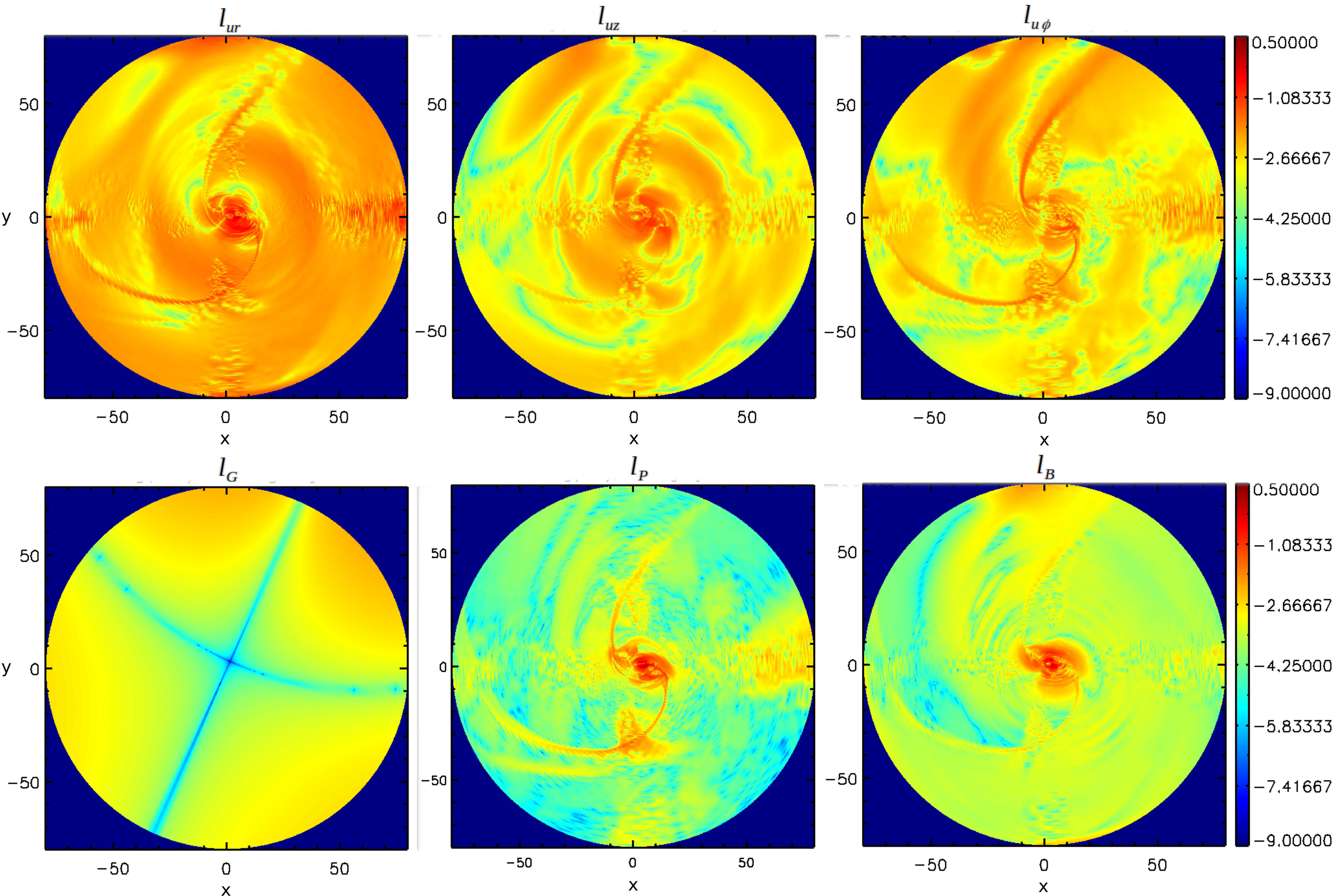}
\caption{Specific angular momentum distribution in the jet at time $t=1500$. 
The presentation is similar to Figure~\ref{fig:eq16_disk}, however the variables displayed are now shown averaged 
from $z=25$ to $z=35$ in the jet flow.
The reason is their pixelized distribution that is resulting from the application of the gradients involved which are calculated in post-processing.}
\label{fig:eq16_jet}
\end{figure*}

Naturally, we would expect that the local launching conditions for the disk wind will be reflected in the disk structure and the
dynamics of the disk wind -- in this case implying that the wind or jet structure also carries a density wave, respectively a spiral arm.
This exactly demonstrated by our simulations.
While the effect might not come as a surprise, its impact on the disk and jet dynamics has never been discussed before.

Obviously, a spiral wave structure of the outflow will have severe implications concerning its stability and 
also the overall angular momentum budget of the disk-jet system (see next section).
We first discuss the overall structure of the outflow launched from the disk.

Figure~\ref{fig:vtor1500} and \ref{fig:vr1500} display snapshots of the radial velocity, the vertical velocity and the 
rotational velocity at $z\simeq 25$.
Again, we see a spiral structure as the prominent features in the velocity pattern which is now inside the jet.
The location of the spiral arms is consistent with the position of the density peaks discussed above.

While we usually discuss the structure of a spiral {\em arm} in the disk, the corresponding structure along the jet is that 
of a wall (of increased density), to which we will refer in the following as a {\em spiral wall}.
When there is differential pattern speed of the jet spiral structure {\em along} the jet, that spiral wall will follow
a {\em helical} structure (see discussion below).

As the disk spiral arms become denser in time, the same happens with the jet spiral walls which are as well rotating 
synchronized with the orbital motion of the binary.

A 3D visualization of the evolution of the spiral features along the disk-jet structure is shown in Fig.~\ref{fig:3dview_box_diffz}
where display consecutive 2D slices along the 3D disk-jet system. 
In particular, the plot visualizes the lag between the spiral features from jet layer to jet layer, which is 
directly connected to the propagation of the jet material.
This strongly supports our claim that the spiral structure is {\it injected} into the outflow from the disk.

We now quantify the spiral wall structure in the outflow and compare it to the disk spiral arms.
We thus focus on the angular profiles of density and rotational velocity at $r=20$ inside the jet, here at $z=25$ 
(see Fig.~\ref{fig:vpatt-jetz30} and compare it to the same plots for the disk area i.e., Fig.~\ref{fig:vpatt}).
We find that the position of the minimum in the mass density of the disk spiral arm area and those of the jet spiral wall area are very close.

The same holds when comparing the angular profiles of the rotation velocity $u_\phi(\phi)$.
Essentially, the spiral features in the disk and in the jet follow the same kind of time evolution, meaning that the 
jet spiral arms do not lag the disk spiral arms, and establishing an almost stationary 
\footnote{Our simulations do not reach a steady state because of two reasons.
Firstly, because of the limited simulation time.
Secondly, and more physically, due to the fact that the disk mass decreases over time due to accretion and ejection.} 
structure, such that the spiral features in the jet are almost co-rotating with the disk spiral features
(as discussed above).

Roughly speaking, the small change in the position angle of the spiral arm along the wall arises from the fact that the jet
dynamical time scale is much faster than the disk dynamical time scale.
As in any accretion-ejection scenario \citep{2004ApJ...601...90C, 2007A&A...469..811Z, 2012ApJ...757...65S} the dynamical 
time scale of the jet is basically defined by the Alfv\'enic 
time scale, and is much faster than the dynamical time scale of the disk which follows the dissipative time scales.

The jet Alfv\'enic time scale is $\tau_{\rm A} = \Delta L / v_{\rm A}$,
where we consider for $\Delta L \simeq 10$, that is either the jet length close to the disk or the jet width. 
At this point the jet is trans-sonic, $v_{\rm A} \simeq 0.125$, thus $\tau_{\rm A} \simeq =10/0.125= 80$.
This is the time scale on which the internal jet structure can be causally changed on these scales.
This time scale is similar (given the trans-sonic jet nature) also to the kinematic time scale 
of the $\tau_{\rm kin} \equiv \Delta L / v_{\rm jet} \simeq =25/0.2=125$ that is the time scale typical for jet propagation.

Launching of the jet out of the disk, i.e the mass transfer from accretion into ejection, happens on a
resistive time scale $\tau_{\eta} \simeq (\Delta L)^2 / \eta$.
Assuming $\Delta L \simeq 1$ for the launching area and an average magnetic diffusivity $\eta \simeq 0.03$, the time scale for 
the launching process is about $\tau_{\eta} \simeq 33$, thus shorter than the time scale we observe for spiral arm formation.

The jet mass flux is certainly fed by the accretion disk.
The feeding is - essentially - a local process, such that each surface area of the disk feeds the outflow 
that is launched from there.
Once injected, the material is rapidly accelerated along the outflow, and any imprint of the injection
process is propagated to larger altitudes.
However, the propagation time of the outflow, together with the spiral arm orbital time scale will
lead to a lag between the spiral arm structure in the disk and in the outflow.

In summary, any structure that develops in the disk, is {"}immediately{"} propagated along the wind.
Nevertheless, on very large spatial scales we would expect the jet spiral arms to lag behind the disk spiral 
arms, assuming that the spiral structure and the jet survives that long.

We find that the vertical jet speed (measured at $z=25$, see Fig.~\ref{fig:vr1500}) is more 
smoothly distributed than the poloidal velocity field and the density in the disk mid plane.
In contrast, the toroidal velocity in the disk is smoother than that for the jet. This is also seen in the angular profiles in 
Fig.~\ref{fig:vpatt}.
We explain this by the observation that at the position of the density spiral arms, also the poloidal magnetic field is enhanced 
(accumulated).
This can be seen in Fig.~\ref{fig:brbphibzjetdisk}, that shows a distinct spiral pattern in the disk for all three field components.
Inside the jet, the spiral features are also present, but are much broader. 
Thus, for launching and accelerating the higher mass flux out of the spiral arm to similar
speed, a stronger magnetic flux is available.

Comparing the mass density distribution in the jet and the counter jet, we find a similar 
profile for the jet spiral arms which just reflects the bipolar symmetry of the setup, 
in particular our model setup with the binary orbit being co-planar with the mid-plane of the disk forming jet.
If the secondary would be instead placed offset to the mid-plane (and the initial disk) establishing 
a bipolar asymmetry in the gravitational potential, we would expect a different spiral structure
for jet and counter jet.

In this section, we have proposed the scenario that the jet spiral structure we observe is injected from the
disk into the outflow and then propagates along the jet.
A valid objection may be that the tidal forces by the binary system that modify the disk structure,
do also act directly on the jet.
In particular for low altitudes, thus the disk wind close to the disk, the tidal forces on disk and jet are expected to be
similar \footnote{Note that the extension of the Roche lobe in vertical direction is about $z\simeq 70$}.
Therefore, the jet flow through the Roche lobe can be tidally deformed, and the resulting structure is
 {\em not} the result of the injection process alone.

However, we have further proven our hypothesis of a disk spiral structure injected into the outflow by a simple 
numerical experiment.
We have first run a jet launching simulation for a single star
 ( see Figure ~\ref{sin_to_bin_spiralwall} in the appendix).
An almost axisymmetric 3D jet structure evolves that reaches a quasi steady-state at time $t\simeq 4000$.
After that time, we then switch on the gravity of the Roche potential and observe how the disk and jet structure 
further evolves.
What we see is that the spiral structures in the disk and in the jet do not arise at the same time.
Instead, the disk spiral structure evolves first, after about 500 further time steps.
Then, subsequently, the different jet layers also exhibit spiral arms.
Furthermore, the jet spiral structure is first seen at lower altitudes (we have compared layers $z=25, 35, 45$),
the at the higher layers.
From layer to layer separated by $\Delta z =20$ it takes the spiral structure another time period of $\Delta t \simeq 100$
to appear. 
This compares to a pattern speed propagating along the jet with $v_{\rm pat} \simeq 0.2$ which is indeed
comparable to the wind or jet speed at this location.
We see this as very strong support for our hypothesis.

Would the spiral structure in disk and jet be produced instantaneously by the tidal forces, such
a time lag would not be visible (see our discussion above).

It would be interesting to follow this features to much higher altitudes, even beyond the Roche lobe in 
vertical direction.
If the jet at these distances would carry a spiral structure, we expect that to be injected from the lower
altitudes, in fact by the disk, as the tidal forces beyond the Roche lobe are considerably smaller.
The gravitational potential far beyond the Roche lobe approaches that of a point source, although we
expect some jet bending towards the secondary not far from the Roche lobe.

\section{Local torque analysis in the disk and the outflow}
In this section, we analyze the forces and torques acting on the disk and the jet in order to investigate 
the physics of 3D effects in MHD jet launching, that is the interaction between the disk and the jet.
Similar works have recently been published, studying hydrodynamic torques in circum-binary disks
\citep{2017MNRAS.466.1170M,2019ApJ...875...66M}, or the torques exerted on accreting supermassive
black hole binaries \citep{2013MNRAS.435.2633N}.
Only a few studies exist that have considered {\em magnetic torques} in a binary system.
One example is \citet{2017MNRAS.469.4258T} who have studied the magnetic torque in accretion disks of 
millisecond pulsars.

A classic problem of MHD winds and jets is the loss of the disk angular momentum by the magnetic torque of 
a disk outflow (see e.g. \citealt{1992ApJ...394..117P}).
It has become clear that MHD disks wind do remove angular momentum from the disk very efficiently, due to extended lever arm.

For the kind of studies just cited above, certain approximations can be made when calculating the angular 
momentum balance.
For example, the initial studies of MHD jet formation were considering steady-state MHD and axisymmetry.
Thus, certain terms could be neglected in the angular momentum balance. 
The same is true for studies concentrating on the disk physics only. 
Here, the vertical motion (in the disk) can be neglected, thus also the vertical angular momentum loss.
This is obviously also the case for purely hydrodynamic studies that do not consider any magnetic torque 
and its accompanied extended lever arm. 

As the present paper considers MHD launching in 3D, we cannot simplify the angular momentum equation accordingly.
Instead, we have to deal with all terms -- the magnetic terms, the angular momentum loss induced by vertical transport ($u_{\rm p} \simeq u_{\mathrm K}$), and, essentially, also the derivatives in the $\phi$-direction.

In order to study the different torques acting on the accretion disk, we concentrate on the toroidal component of the momentum defined by Equation~\ref{momentum_eq} and the $\phi$-component is given as
\begin{multline}
\frac{\partial u_{\phi}}{\partial t}+\frac{u_{r}}{r} \frac{\partial (r u_{\phi})}{\partial r}
+\frac{u_{z}}{r} \frac{\partial (r u_{\phi})}{\partial z} 
+\frac{u_{\phi}}{r} \frac{\partial u_{\phi}}{\partial \phi}=\\
-\frac{1}{r}\frac{\partial \Phi}{\partial \phi}
-\frac{1}{\rho r}\frac{\partial P}{\partial \phi}\\
+\frac{1}{4\pi \rho r}
 \left[ B_{r} \frac{\partial (r B_{\phi})}{\partial r} 
      + B_{z} \frac{\partial (r B_{\phi})}{\partial z}
      - \frac{\partial \left(B_r ^2 + B_z ^2\right) }{2 \partial \phi}.
 \right]
\label{phi_momentum_eq0}
 \end{multline}
We can re-write this equation as
  \begin{multline}
  \frac{\partial u_{\phi}}{\partial t}
+ \frac{1}{r} \vec{u}_P\cdot\nabla(r u_{\phi})
+ \frac{u_{\phi}}{r} \frac{\partial u_{\phi}}{\partial \phi}=\\
-\frac{1}{r}\frac{\partial \Phi}{\partial \phi}
- \frac{1}{\rho r}\frac{\partial P}{\partial \phi}\\
+ \frac{1}{4\pi\rho r} \left[\vec{B}_P\cdot\nabla(r B_{\phi})
- \frac{\partial \left(B_r ^2 + B_z ^2\right) }{2\partial \phi} \right].
\label{phi_momcf1}
 \end{multline}
It is immediately clear that there are terms in this fully 3D equation which are related to derivatives 
in $\phi$-direction.
When applying stationarity, ${\partial }/{\partial t}=0 $, and axisymmetry, ${\partial }/{\partial \phi}=0 $,
and using $\vec \nabla \cdot \rho \vec u_{\rm p} =0$ and $\vec \nabla \cdot \vec B_{\rm p} =0$,
we may obtain another, simplified form of equation \ref{phi_momcf1},
 \begin{equation}
 \vec \nabla \cdot \left( \vec{u}_{\rm p} ru_{\phi}-\vec{B}_{\rm p} r u_{\phi} \right)=\vec \nabla\cdot \vec{\tau}_{0},
 \end{equation}
 
where $\vec{\tau}_{0}$ represents the other torques acting in the system, such as e.g. the viscous torque.
By applying Stokes theorem, this equation can be converted to
\begin{equation}
    \tau_{0} = \int_S r \left(\rho u_{\phi} \vec{u}_p - \frac{1}{4\pi} B_\phi \vec{B}_p \right) \cdot \vec ds.
    \label{eq-tau-PP92}
\end{equation}
which is the well-known equation for the angular momentum flux (thus, the torques) 
derived for steady-state MHD wind theory (see e.g. \citealt{1992ApJ...394..117P, 2013A&A...550A..99Z}).
As for another example -- that also connects to the topical literature -- we may consider non-stationarity,
but focus on the angular momentum budget only inside the disk. 
This approach is typical when investigating solely the hydrodynamic disk structure 
(see e.g. \citealt{2017MNRAS.466.1170M, 2019ApJ...875...66M}).
Neglecting the vertical motion in the disk, $u_z \simeq 0$, in Equation~\ref{phi_momentum_eq0},
and integrating vertically over the disk, we arrive at
\begin{multline}
   \frac{\partial u_{\phi}}{\partial t}
 + \frac{u_{r}}{r}    \frac{\partial \left( r u_{\phi} \right) }{\partial r}
 + \frac{u_\phi u_r}{r}
 + \frac{u_{\phi}}{r} \frac{\partial u_{\phi}}{\partial \phi} = \\
 - \frac{1}{r} \frac{\partial \Phi}{\partial \phi}
 - \frac{1}{\Sigma r} \frac{\partial P}{\partial \phi}
 + \int \frac{1}{\rho} F_{\rm B} dz
\label{phi_momentum_equ1}
 \end{multline}
which is similar to Equation~A2 in \citet{2017MNRAS.466.1170M}.
Here $\Sigma$ denotes the surface density of the disk, while the Lorentz force $F_{\rm B}$ in the $\phi$-component of the equation of motion
is defined as
\begin{multline}
F_{\rm B} = \frac{1}{4\pi r} \left[
      B_{r} \frac{\partial \left(r B_{\phi} \right) }{\partial r} + 
      B_{z} \frac{\partial \left(r B_{\phi} \right) }{\partial z} - 
      \frac{\partial \left(B_r^2 + B_z^2 \right) }{2\partial \phi}
                             \right],
      \label{eq:tau-bx}
\end{multline}
In our approach that considers disk outflows and jets, we cannot ignore the vertical motion.
We thus need to consider the full Equation~\ref{phi_momentum_eq0} 
that allows us to study the angular momentum budget inside disk and along the jet.
Multiplying Equation~\ref{phi_momentum_eq0} by $r$ we find for the evolution of the specific angular momentum $l = r u_{\phi}$,
\begin{multline}
  \frac{\partial l}{\partial t}=
 -       u_r        \frac{\partial l}{\partial r} 
 -       u_z        \frac{\partial l}{\partial z}
 - \frac{u_\phi}{r} \frac{\partial l}{\partial \phi} \\
 - \frac{\partial \Phi}{\partial \phi}
 - \frac{1}{\rho}   \frac{\partial P}{\partial \phi} 
 + \frac{1}{\rho}  r F_{\rm B}. 
\label{llintime}
\end{multline}
On the r.h.s.~of Equation~\ref{llintime} different torques 
appear that affect the disk specific angular momentum evolution - 
namely the pressure gradient torque, 
the gravity torque and the magnetic torque $\tau_B = r F_{\rm B}$, respectively.
Here, also the $\phi$-derivative of the magnetic pressure term is included that has been neglected for Equation~\ref{eq-tau-PP92}.
\begin{figure*}
\centering
\includegraphics[width=18cm]{\figurepath/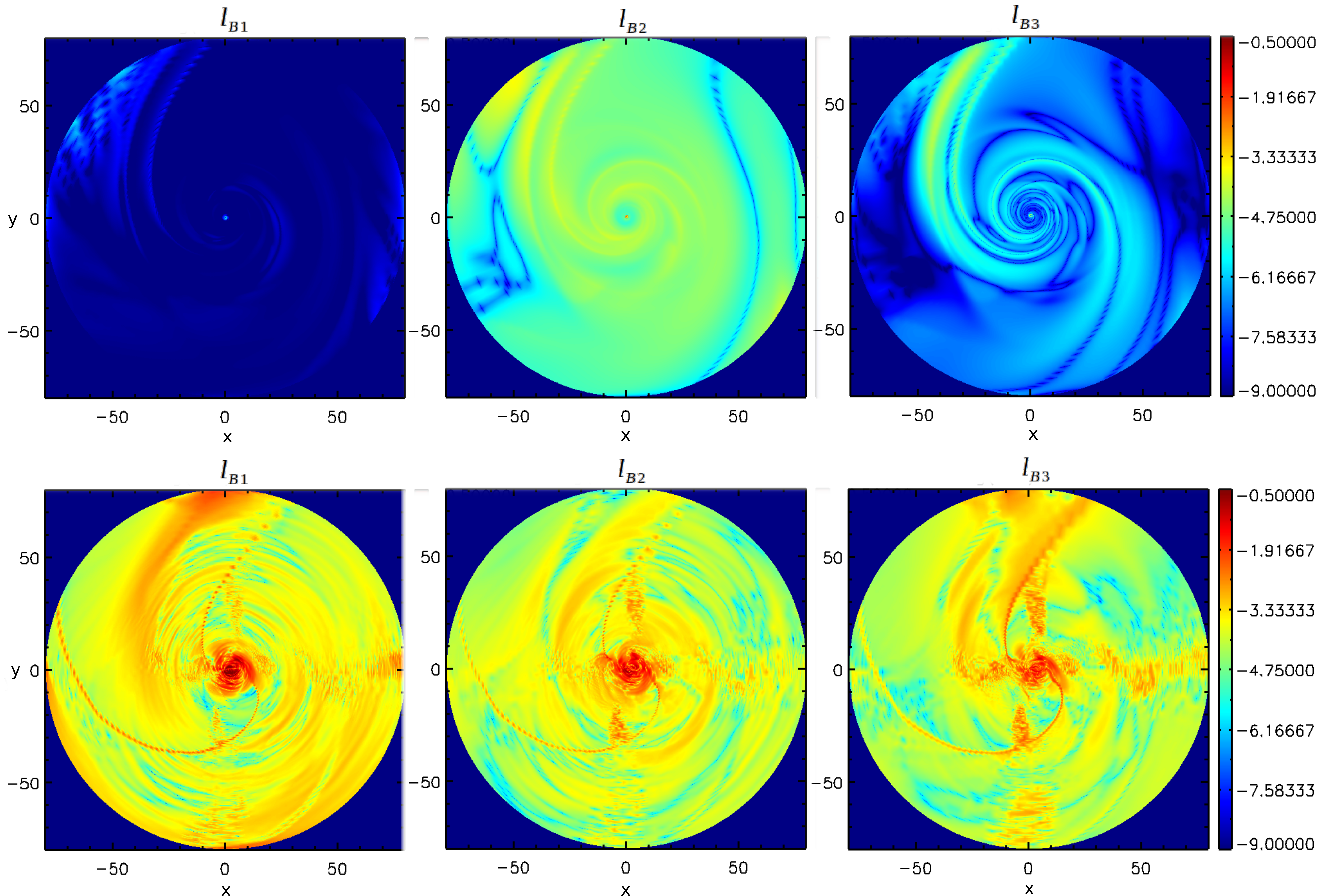}
\caption{Specific magnetic angular momentum contributions in equation \ref{llintime} at time $t=1500$.
These terms are shown along the disk mid-plane (top) and in the jet (bottom) at $z=25$.
For comparison these terms are listed in
Table~\ref{tbl:terms} (top).}
\label{tb_term_magnetic}
\end{figure*}
We now compare the different contributions in Equation~\ref{llintime}. For convenience, we display them in Table~\ref{tbl:terms} (top) in consecutive order.
In Figures~\ref{fig:eq16_disk} and \ref{fig:eq16_jet} we show these terms at time t=1500.
The main conclusions from our comparison are the following.

(i) Among all terms, the gravity torque $l_{\rm G}$ (see Table~\ref{tbl:terms}, top) is most smoothly 
distributed\footnote{We stress again that the pixelized appearance of some of the terms arises
from the numerical calculation of the gradients involved.
That is, however, obtained in post-processing (in cylindrical coordinates, interpolated from the numerical Cartesian grid), so the simulation procedure itself is not affected.}.
This is easy to understand as this term does not depend on the MHD variables like density, but it is just defined by the time evolution of the Roche potential, i.e. the location of the companion star.
Consequently, we observe that the signature of the spiral arms is clearly visible in all panels except the gravity torque.\\

(ii) The term $l_{\rm Uz}$  corresponding to the exchange of specific angular momentum due to vertical transport is larger 
inside the jet compared to the disk. 
This may simply be explained by the fact that the vertical advection speed is much smaller in the disk compared to the jet.\\ 

(iii) There are more spiral windings in the disk than for the jet.
We understand this as a consequence of the opening-up of the jet flow.
Due to the opening cone of the outflow the opening angle of a spiral wave injected into the outflow decreases 
with altitude and eventually the wave pattern dies out.\\

(iv) The spiral arms in the disk and inside the jet are not synchronized. 
The spiral arm in the jet lags the spiral arm in the disk (best visible for $l_{\rm P}$). 
We understand this as due to the jet inertia. 
The jet is set in rotation by the magnetic field that is anchored in the disk
(like a whirlpool).
The inertia of the jet material counteracts the toroidal Lorentz force and leads to a lag between the foot point 
of the jet and the jet upper layers.
We also note that the jet rotation pattern we observe at a certain time results from an injection from the disk 
into the jet at earlier times (when the disk spiral structure was located at an earlier position). 
This pattern is then propagated to higher altitudes.
A simple estimate of the time lag is by comparing the (estimated) outflow speed and the altitude in the jet 
$\Delta t = z / u_{\rm jet} \simeq 25 / 0.5 = 50$
which roughly fits to what we observe in our simulation pattern comparing similar time differences.\\ 

(v) For a comparison of the magnetic terms defined by Equation~\ref{eq:tau-bx}, we look at in Figure\ref{tb_term_magnetic}.
The terms are shown for the disk (mid-plane) and for the jet. 
It is obvious that all three terms are larger in the jet compared to the disk.
This is due to the large magnetic lever arm in the jet and is known from traditional (axisymmetric) jet theory.
Essentially, the figure also demonstrates that the non-axisymmetric effects are crucial even 
far from the disk mid-plane.\\

(vi) The magnetic term also shows a spiral structure. 
This can be explained by the fact that the magnetic field is frozen into the disk material.
Therefore, the magnetic flux follows the disk density structure, thus the spiral shape. 
On the other hand the launching of the spiral structure from the disk into the outflow is affected by the disk resistivity.
So the jet spiral magnetic field may lag the disk spiral.
Once loaded into the disk wind, the further evolution follow ideal MHD.
However, inertial forces (caused by the indirect term in time dependent Roche potential, see Equation~\ref{roche_potential})
will continue to affect the jet spiral structure, with the result that the spiral 
in the more distant parts of the jet lags the spiral in the parts of the outflow close to the disk.\\

In summary, we find that the 3D effects such as the $\phi$-dependency of various variables as well as the vertical transport of the material are essential for our study of the angular momentum budget in the disk-jet system.
The non axisymmetric structures that are triggered in the disk by the Roche gravitational 
potential are launched into the disk wind and are propagated into the jet.
In the next section we will discuss and show how these effects contribute to the 
overall angular momentum budget of the disk and jet.

\section{Global angular momentum balance}
We finally briefly investigate the global angular momentum budget and compare the respective impact 
of the particular physical terms.
We integrate the differential equation discussed above and compare the radial and vertical profiles of the angular momentum distribution over time.

Multiplying Equation~\ref{llintime} by $r\rho$, and making use of both the identity
\begin{equation}
\rho \frac{\partial l}{\partial t}
  = \frac{\partial \left( \rho l \right) }{\partial t} - l\frac{\partial \rho}{\partial t}
\end{equation}
and the continuity equation, we arrive at
\begin{multline}
  \frac{\partial}{\partial t} \left(r \rho l\right)=
 -\frac{\partial}{\partial r} \left(\rho r u_r l\right)
  -\frac{\partial}{\partial z} \left(\rho r u_z l\right)
 -\frac{\partial}{\partial \phi} \left(\rho u_\phi l\right)\\
 - r\rho \frac{\partial \Phi}{\partial \phi}
 -r\frac{\partial P}{\partial \phi}
 +r^2  F_{\rm B}.
 \label{phi-torque1}
\end{multline}
Equation~\ref{phi-torque1} is identical to Equation~(A6) in \citet{2017MNRAS.466.1170M},
except for the two additional terms that appear in our approach and are due to the existence of 
the (i) magnetic field and the (ii) disk wind. 

In order to calculate the poloidal angular momentum fluxes (i.e. accretion and ejection), 
respectively the profile of the torques at work,
we need to integrate Equation~\ref{phi-torque1} in the corresponding poloidal directions.
We decided to integrate (i) in vertical direction ($z$-direction) and (ii) in radial direction.
This will deliver the angular momentum flux that is advected (i) along the disk and (ii) into the 
outflow.

\subsection{Radial angular momentum balance}
By integrating Equation~\ref{phi-torque1} in $\phi$ and $z$-direction, we arrive at

\begin{multline}
  \frac{\partial}{\partial t}\int \int  \left(r \rho l\right) \,d\phi\,dz
 = -\int \int \frac{\partial}{\partial r} \left(\rho r u_r l\right) \,d\phi\,dz \\
   -\int \int \frac{\partial}{\partial z} \left(\rho r u_z l\right) \,d\,\phi\,dz
   -\int \int \frac{\partial}{\partial \phi} \left(\rho u_\phi l\right) \,d\phi\,dz\\
   - \int \int  r\rho \frac{\partial \Phi}{\partial \phi} \,d\phi\,dz
   -\int \int r\frac{\partial P}{\partial \phi} \,d\phi\,dz
   +\int \int r^2 F_{\rm B} \,d\phi\,dz.
 \label{phi-torque2}
\end{multline}
The integrated values may also be understood as averages over the $z$ and $\phi$-direction.
The integration area is chosen as one (initial) disk scale height in vertical direction, 
$\Delta z (r) = [-0.1~r, 0.1~r]$, and $\Delta \phi=[0, 2\pi]$, 
thus confined to the disk region.

\begin{figure*}
 \centering
\includegraphics[width=18cm]{\figurepath/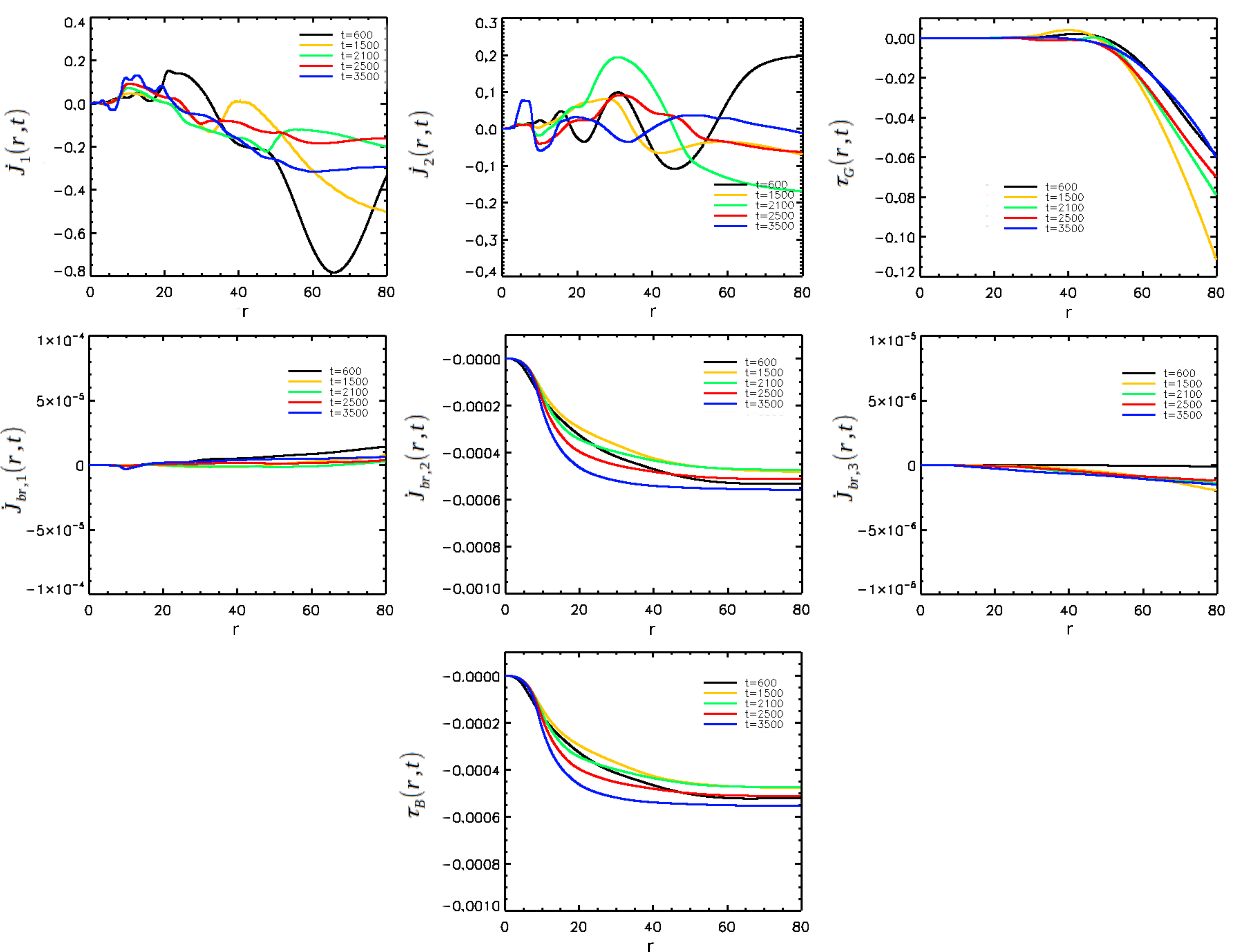} 
\caption{Angular momentum flux evolution for a binary disk-jet system.
Shown are radial profiles of the different contributions to the total angular momentum 
flux $\dot{J}(r,t)$ for different times.
These terms are 
$\dot{J}_1(r,t)$ considering the radial advection of angular momentum,
$\dot{J}_2(r,t)$ considering the the vertical transport, 
$\tau_{\rm G}(r,t)$ considering the gravity torque, 
and $\tau_{\rm B}(r,t)$ the magnetic torque.
The terms $\dot{J}_{\rm br,1}(r,t)$, $\dot{J}_{\rm br,2}(r,t)$ and $\dot{J}_{\rm br,3}(r,t)$ represent the different contributions 
to the magnetic torque, respectively (see Table~\ref{tbl:terms}).
For each radius $r$ these terms are integrated from $r=0$ to $r$, while we have vertically averaged between
$z=-0.1~r$ and $z=0.1~r$.
The terms $\dot{J}_3(r,t)$ for the orbital transport is quite small and $\tau_{\rm P}(r,t)$ for the pressure torque
approximately vanishes and are not shown here.}
\label{jdotr_averged_binary}
\end{figure*}

\begin{figure*}
 \centering
\includegraphics[width=18cm]{\figurepath/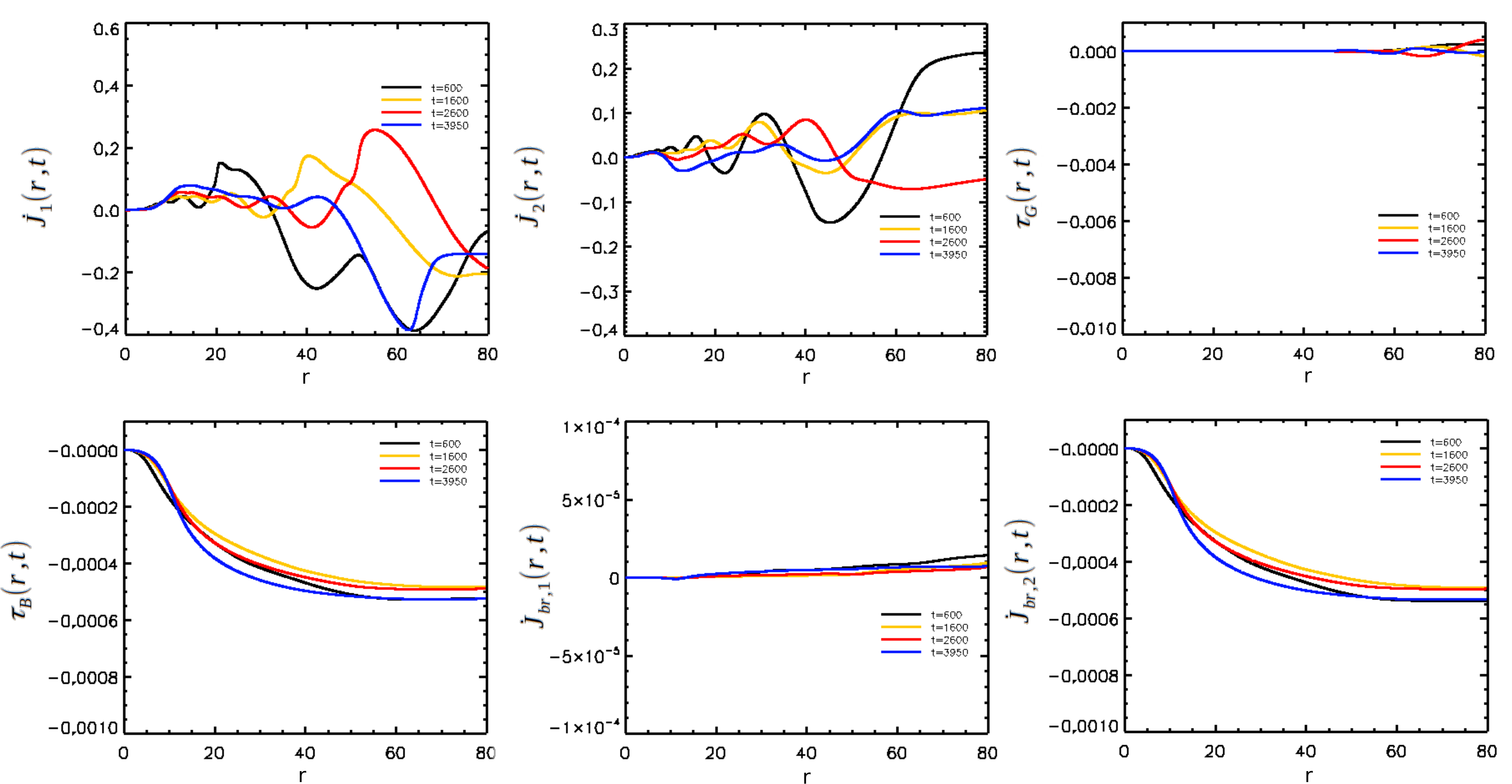} 
\caption{Angular momentum flux evolution for a single star disk-jet system.
For comparison with Figure~\ref{jdotr_averged_binary}, we have reproduced the same terms for a single star jet launching simulation. 
Shown are radial profiles of the different contributions to the total angular momentum 
flux $\dot{J}(r,t)$ at different times.
These terms are 
$\dot{J}_1(r,t)$ which considers the radial advection of angular momentum,
$\dot{J}_2(r,t)$ considering the the vertical transport, 
$\tau_{\rm G}(r,t)$ considering the gravity torque
and $\tau_{\rm B}(r,t)$ the magnetic torque.
The terms $\dot{J}_{\rm br,1}(r,t)$, $\dot{J}_{\rm br,2}(r,t)$ and $\dot{J}_{\rm br,3}(r,t)$ represent the different contributions 
to the magnetic torque, respectively (see Table~\ref{tbl:terms}).
For each radius $r$ these terms are integrated from $r=0$ to $r$ while we have vertically averaged between
$z=-0.1~r$ and $z=0.1~r$. 
The terms $\dot{J}_3(r,t)$ for the orbital transport and $\tau_{\rm P}(r,t)$ the pressure torque, and 
$\dot{J}_{\rm br,3}(r,t)$ do vanish and are not shown here.}
\label{jdotr_averged_single}
\end{figure*}

The different terms in Equation~\ref{phi-torque2} contribute differently to the overall angular momentum 
evolution of the binary star-disk-jet system.
Altogether, Equation~\ref{phi-torque2} describes the rate of change of angular momentum across a ring with 
radius $r$ and width $dr$.

When integrating $\rho l$ over the volume element $dz\,r d\phi$, we obtain the total angular momentum in a cylinder of
width $dr$ and radius $r$ (see left hand side of Equation~\ref{phi-torque2}). 
Equation~\ref{phi-torque2} can be re-written as
 \begin{multline}
\frac{\partial }{\partial t}\left( \frac{dJ_{\rm tot}(r,t)}{dr} \right)=\\
 \frac{\partial}{\partial r} \left( \dot J_1(r,t)\right)+\frac{\partial}{\partial r}\left( \dot J_2(r,t)\right)
 +\frac{\partial }{\partial r} \left( \dot J_3(r,t)\right)\\
 +\frac{\partial\tau_{\rm G}(r,t)}{\partial r} +\frac{\partial\tau_{\rm P}(r,t)}{\partial r} 
  +\frac{\partial  \tau_{\rm B}(r,t)}{\partial r},
  \label{allJr}
\end{multline}
where we define the following terms contributing to the total angular momentum evolution
\begin{equation}
 \frac{d \dot{J}_{\rm tot}(r,t)}{dr} = \frac{\partial}{\partial t} \oint \int r \rho l \, dz \,d\phi.
 \label{djdr}
\end{equation}
These contributions are the inward flux of angular momentum due to radial advection,
\begin{equation}
 \frac{d \dot{J}_{1}(r,t)}{dr} 
   = -\oint \int \frac{\partial}{\partial r} \left(\rho r u_r l \right)\, dz \, d\phi,
 \label{j1_adv}
\end{equation}
the loss of angular momentum due to vertical transport,
\begin{equation}
 \frac{d \dot{J}_{2}(r,t)}{dr} =
 -\int \int \frac{\partial}{\partial z} \left( \rho r u_z l \right)\, d\,\phi\,  dz,
 \label{j2_up}
\end{equation}
the radial flux of angular momentum due to toroidal motion,
\begin{equation}
 \frac{d \dot{J}_{3}(r,t)}{dr} =
  -\int \int \frac{\partial}{\partial \phi} \left( \rho u_\phi l \right) \,d\phi \,  dz,
 \label{j3_shear}
\end{equation}
the gravitational torque per unit radius, 
\begin{equation}
 \frac{d\tau_{\rm G}(r,t)}{dr} = -\oint \int r\rho \, dz \frac{\partial \Phi}{\partial \phi} \, d\phi,
\label{grtor}
\end{equation}
the pressure torque per unit radius,
\begin{equation}
 \frac{d\tau_{\rm P}(r,t)}{dr} =-\int \int r\frac{\partial P}{\partial \phi} d\phi  dz,
\label{gpre}
\end{equation}
and the magnetic torque per unit radius.
\begin{equation}
 \frac{\partial \tau_{\rm B}(r,t)}{\partial r}=\oint \int r^2 F_{\rm B}\,dz\, d\phi.
\label{Jmag}
\end{equation}
Since we conduct a fully 3D study, a few extra terms are seen compared to \citet{2017MNRAS.466.1170M}.
These are torques due to vertical and azimuthal motion or the thermal pressure torque or magnetic torque.
Among the different terms, the disk wind plays a major role in the angular momentum transport.
More specifically, the Equations~\ref{j2_up} and \ref{Jmag} provide the contribution of the disk outflow 
for the angular momentum budget of the system.

We notice that Equations~\ref{djdr}-\ref{Jmag} describe $\partial_r \dot J(r,t)$ or $\partial_r \tau(r,t)$. 
Thus, to derive the radial profile of the angular momentum fluxes at work, respectively the torques,
we integrate each term  
(see Table~\ref{tbl:terms}, middle) 
in $r$-direction from $0$ to $r$.
For example, to obtain
$\dot{J}_1(r,t)$ at each radius point we integrate
$\dot{J}_1(r,t) = \int_0^r \, Te_{\rm r,1}(r',t) dr'$ and $Te_{\rm r,1}$ defined in Table~\ref{tbl:terms}. 
We notice that $\dot{J}_1(r,t) $ represents the angular momentum flux of radial transport, integrated in $\phi$ and $z$-direction, and also along the radial direction 
at each radius.

As another example, we can consider the different terms of the radial magnetic torque which are obtained as 
in the following example $ {\tau}_{B}$.
\begin{equation}
\tau_{\rm B}(r,t)= \dot{J}_{br,1}(r,t) + \dot{J}_{br,2}(r,t)+ \dot{J}_{br,3}(r,t)
\label{jdot6_terms}
\end{equation}
with
\begin{equation}
 \dot{J}_{br,1}(r,t)=\int_0^r Te_{\rm B,r,1}(r',t) dr'
\end{equation} 
and the terms $Te_{\rm B,r,1}, Te_{\rm B,r,2},Te_{\rm B,r,3} $ are defined in Table~\ref{tbl:terms}.

In Figure~\ref{jdotr_averged_binary} we display the radial profile of the angular momentum fluxes and the 
corresponding torques on the disk-jet at different times 
(see again Table~\ref{tbl:terms}, middle).

In order to disentangle the contribution of the 3D terms to the local and global angular momentum budget, 
it is essential to compare the results for the binary star simulation to that of a single star.
We thus show the respective terms also for a single star simulation in Figure~\ref{jdotr_averged_single}.
This simulations is run in fully 3D \citep{2015ApJ...814..113S}  but with single star gravitational potential (thus no time-dependent Roche potential). 
We compare in detail all terms that contribute to the radial profile of the angular momentum transport 
in the disk by different physical processes, respectively (see Table~\ref{tbl:terms}, middle).
As we find some of the terms being negligibly small, we do not show them in the Figure~\ref{jdotr_averged_single}.
By comparing Figure~\ref{jdotr_averged_binary} and Figure~\ref{jdotr_averged_single} we derive the 
following conclusions.\\

(i) Here, we study the angular momentum evolution of a MHD accretion-ejection structure 
orbiting in a binary system, thus our approach is magnetized and non-axisymmetric. 
Compared to previous works 
\citep{1977MNRAS.181..441P, 1979MNRAS.186..799L,2017MNRAS.468.1387L,2020A&A...635A.204A,2020A&A...641A..64H} 
studying the torques acting in a binary system (mostly performed in hydrodynamic limit),
our simulations consider the full magnetic torque, 
thus, all terms for the magnetic tension and the magnetic pressure are taken into account.
In addition, due to the presence of the outflow, the vertical distribution of angular momentum, 
thus the vertical transport of angular momentum, is considered in our approach.\\

(ii) Comparing  Figure~\ref{jdotr_averged_binary} and Figure~\ref{jdotr_averged_single} we see that the $\phi$-dependent 
terms 
$\tau_{\rm G}(r,t)$ and $ \dot J_{\rm br3}$ (part of the magnetic torque), 
are  actually quite contributing
in the binary setup. 
These terms become important due to the orbiting companion star that breaks the axial symmetry, and, thus, represent the {\em 3D tidal effects} in our approach.\\

(iii) Figures~\ref{jdotr_averged_binary} and \ref{jdotr_averged_single} 
present the time evolution of the angular momentum or torques. 
Thus, the sign of the term considered determines if this particular region of the disk is losing or gaining
angular momentum.
Regarding this, we observe that the gravity torques ${\tau}_{\rm G}(r,t)$ and the magnetic torque $\tau_{\rm B}(r,t)$ are 
reducing (removing) the angular momentum through the whole disk area, 
especially at the outer part of the disk. 
In the other panels, corresponding to the terms for advection or vertical transport, the sign varies 
(in time and radius).
Thus, these terms contribute in increasing or decreasing the angular momentum in a particular region, respectively.\\

(iv) We recognize 
for both the single star and the binary approach that the torque 
carried by vertical motion $\dot{J}_2(r,t)$ is comparable to the other terms, such as e.g. the advection torque.
At time $t=2500$ we find  that the profile of $\dot{J}_2(r,t)$ is more scattered, and also larger in binary system.
The reasons are a larger gradient for the vertical velocity and also
stronger velocity fluctuations 
along disk mid-plane for the case of the binary simulation (The figure not shown here). 
We conclude, that the vertical transport considered in the disk has a significant effect. 
This seems to be caused mainly by the existence of an outflow.\\

(v) An essential torque that needs to be considered is that of gravity $\tau_{\rm G}(r,t)$. 
This is a 3D effect and caused by the tidal forces produced by the time-dependent Roche potential.
These features are (obviously) not seen in the gravity torque of the single star 
(see Figure~\ref{jdotr_averged_single}).
In the end, it is this tidal torque that is the fundamental cause for the other 3D effects appearing
in our disk-jet system, including the disk spiral arms (seen in density but also the magnetic field) and in the outflow.
Considering the radial profile of the gravity torque, we observe that it is dominant at the outer part of the disk.
In the inner part ($r<25$), the gravity torque is somewhat smoother but essentially smaller.
The obvious explanation is by means of the gravitational potential. 
At the inner disk region the point gravity of the primary is dominant, while further out the Roche potential plays the dominant role.

(vi) The other substantial term is the magnetic torque $\tau_{\rm B}(r,t)$.
Three different terms are involved in the magnetic torque (see Equation \ref{jdot6_terms} and Table~\ref{tbl:terms}, middle). 
We find that among these terms, the term $\dot{J}_{br,2}$ is dominant in both the binary and the single star setup.
The other terms do not have a serious contribution. 

We see that the magnetic torque $\tau_{\rm B}(r,t)$ is larger in the outer disk regions ($r> 20$). 
This may reflect again the importance of the magnetic lever arm which is larger at the larger radii (less collimated field).
We also see that the first term $\dot{J}_{br,1}$ does not contribute much to the radial angular momentum flux.
This is understandable, as both the $r$-component of the magnetic field is small, and also the
$z$-derivative of the toroidal field is larger than its $r$-derivative. 
Note that along the disk mid-plane the $B_{\phi}$ almost vanishes (in axisymmetric steady state MHD it vanishes
by definition), similar for the component $B_r$.
Thus no contribution due to $B_{\phi}B_r$ stresses is expected here.

Accordingly, we find that the magnetic torques and the gravity torques remove the angular momentum from the disk and support the inward motion of 
the disk material.
However, for the other torques, we do not find a unique behavior throughout the disk.
These torques change their sign at various radial positions.

(vii) Among the different torques we have explored, the torque induced by the pressure gradient is almost 
vanishing and thus do not contribute to the total angular momentum budget.
Also, the torque induced by orbital motion $\dot{J}_3(r,t)$ has quite small contribution to the total angular momentum budget. 
Thus, in agreement with the previous works which do not consider the torque by pressure gradient
\citep{2013MNRAS.435.2633N, 2017MNRAS.469.4258T, 2017MNRAS.466.1170M, 2019ApJ...875...66M},
we can ignore the pressure torque, i.e. the angular momentum transport due to the $\phi$-derivative of the gas pressure.\\

We summarize this subsection by stating again that we have considered in our simulations and in our analysis
the full magnetic torque and also the presence of an outflow, thus angular momentum transport by vertical motion.
After all, among the extra terms considered, this latter term has a significant role on the total angular momentum
budget also in a binary system.
The same holds for the magnetic torque, however, the contribution of the $\phi$-derivative of the magnetic pressure 
and the $B_{\phi}B_r$ stresses are small  in the mid-plane. 

\begin{figure*}
 \centering
\includegraphics[width=18cm]{\figurepath/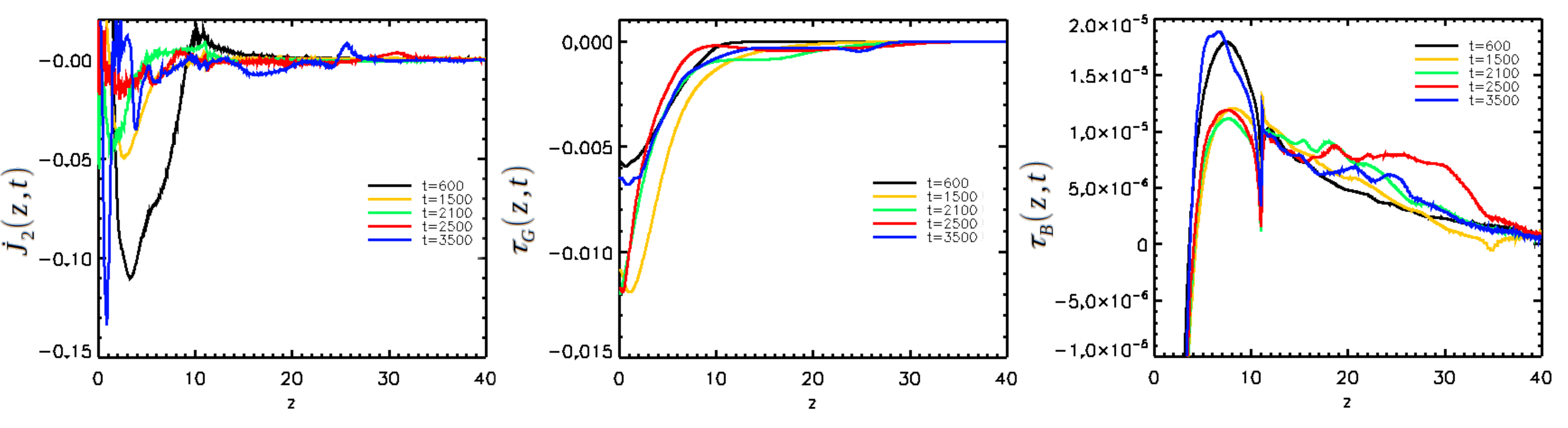} 
\caption{Angular momentum flux along the jet. 
Shown are the vertical profiles of the angular momentum fluxes and torques in the upper hemisphere, $\dot J(z,t)$.
Here, the different angular momentum fluxes are defined by $\dot{J}_2(z,t)$, the vertical transport, $\tau_{\rm G}(z,t)$, 
the gravity torque, and $\tau_{\rm B}(z,t)$, the magnetic torque, respectively.}
\label{jdotz_andt}
\includegraphics[width=18cm]{\figurepath/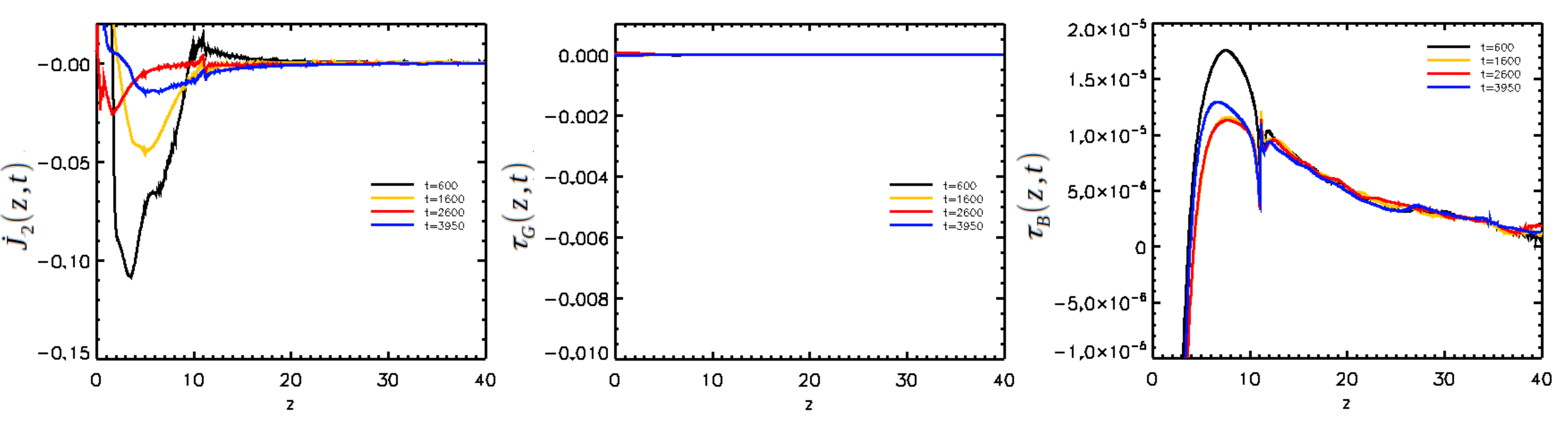} 
\caption{Angular momentum flux along the jet, comparison for a single star simulation. 
Shown are the vertical profiles of the angular momentum fluxes and torques in the upper hemisphere, $\dot J(z,t)$.
Again, the different angular momentum fluxes are defined by $\dot{J}_2(z,t)$, the vertical transport, $\tau_{\rm G}(z,t)$, 
the gravity torque, and $\tau_{\rm B}(z,t)$, the magnetic torque, respectively.
}
\label{jdotz_andtSingle}
\end{figure*}
\subsection{Vertical angular momentum balance}
In the next step, we evaluate the vertical profile of angular momentum transport.
We thus integrate Equation~\ref{phi-torque1} in radial and $\phi$ direction,
\begin{multline}
  \frac{\partial}{\partial t}\int \int  \left(r \rho l\right) \,d\phi \,  dr
 =-\int \int \frac{\partial}{\partial r} \left(\rho r u_r l\right) \,d\phi \, dr \\
 -\int \int \frac{\partial}{\partial z} \left(\rho r u_z l\right)\, d\,\phi\,  dr
 -\int \int \frac{\partial}{\partial \phi} \left(\rho u_\phi l\right) \,d\phi \,  dr\\
 -\int \int r\frac{\partial P}{\partial \phi} d\phi  dr
 - \int \int  r\rho \frac{\partial \Phi}{\partial \phi}\, d\phi \, dr
 +\int \int r^2 F_{\rm B}\, d\phi \, dr.
 \label{phi-torque_3vert}
\end{multline}
For the integration area we have chosen $\Delta r=[0, 80]$ and $\Delta \phi=[0, 2\pi]$. 
Equation~\ref{phi-torque_3vert} describes the rate of change of angular momentum across the surface of a
cylinder with the width of $dz$).
After all, this will provide vertical profiles along the whole disk-jet area.
Equation~\ref{phi-torque_3vert} can be re-written as
\begin{multline}
 \frac{\partial }{\partial t}J_{\rm tot}(z,t) =
 \dot{J}_1(z,t)+ \dot{J}_2(z,t)
 + \dot{J}_3(r,t)\\
 +\tau_{\rm G}(z,t) +\tau_{\rm P}(z,t) +\tau_{\rm B}(z,t)
  \label{allJz}
\end{multline}
By this integration we obtain the vertical angular momentum flux along the whole disk surface.
Similar to the radial profile of the angular momentum (last subsection), 
we define the following terms also for the vertical profile of the angular momentum evolution,
\begin{equation}
 \dot J_{tot}(z,t)= \frac{\partial}{\partial t}\oint \int r \rho l \, dr\,d\phi,
 \label{djdr_z}
\end{equation}
as the vertical flux of angular momentum due to advection,
\begin{equation}
 \dot J_1(z,t) =-\oint \int \rho r u_r l\, dr \, d\phi,
 \label{j1_advz}
\end{equation}
as the vertical flux of angular momentum due to vertical transport,
\begin{equation}
 \dot J_2(z,t) =-\int \int \frac{\partial}{\partial z} \left(\rho r u_z l\right)\, d\,\phi\,  dr,
 \label{j2_upz}
\end{equation}
as the vertical flux of angular momentum due to toroidal motion,
\begin{equation}
 \dot J_3(z,t) =-\int \int \frac{\partial}{\partial \phi} \left(\rho u_\phi l\right) \,d\phi \,  dr
 \label{j3_shearz}
\end{equation}
as the gravitational torque per unit height, 
\begin{equation}
 \tau_{\rm G}(z,t) = -\oint \int r\rho \, dr \frac{\partial \Phi}{\partial \phi} \, d\phi,
\label{grtorz}
\end{equation}
as the pressure torque per unit height,
\begin{equation}
 \tau_{\rm P}(z,t) =-\int \int r\frac{\partial P}{\partial \phi} d\phi  dr
\label{gprez}
\end{equation}
and as the magnetic torque per unit height,
\begin{equation}
 \tau_{\rm B}(z,t)=\oint \int r^2 F_{\rm B}\,dr\, d\phi.
\label{Jmagz}
\end{equation}
In Figure~\ref{jdotz_andt} we display these vertical profiles for the different terms of 
Equation~\ref{phi-torque_3vert} for different times.
These terms correspond to the vertical profile of the angular momentum flux along the jet that is generated by the different 
physical agents in the disk 
(see Table~\ref{tbl:terms}, bottom).
As we find that some of the terms are negligibly small, we have not shown them in the figure.
When comparing the different terms which are presented in Figure~\ref{jdotz_andt} we come to the following conclusions:\\

\begin{figure*}
 \includegraphics[width=18cm]{\figurepath/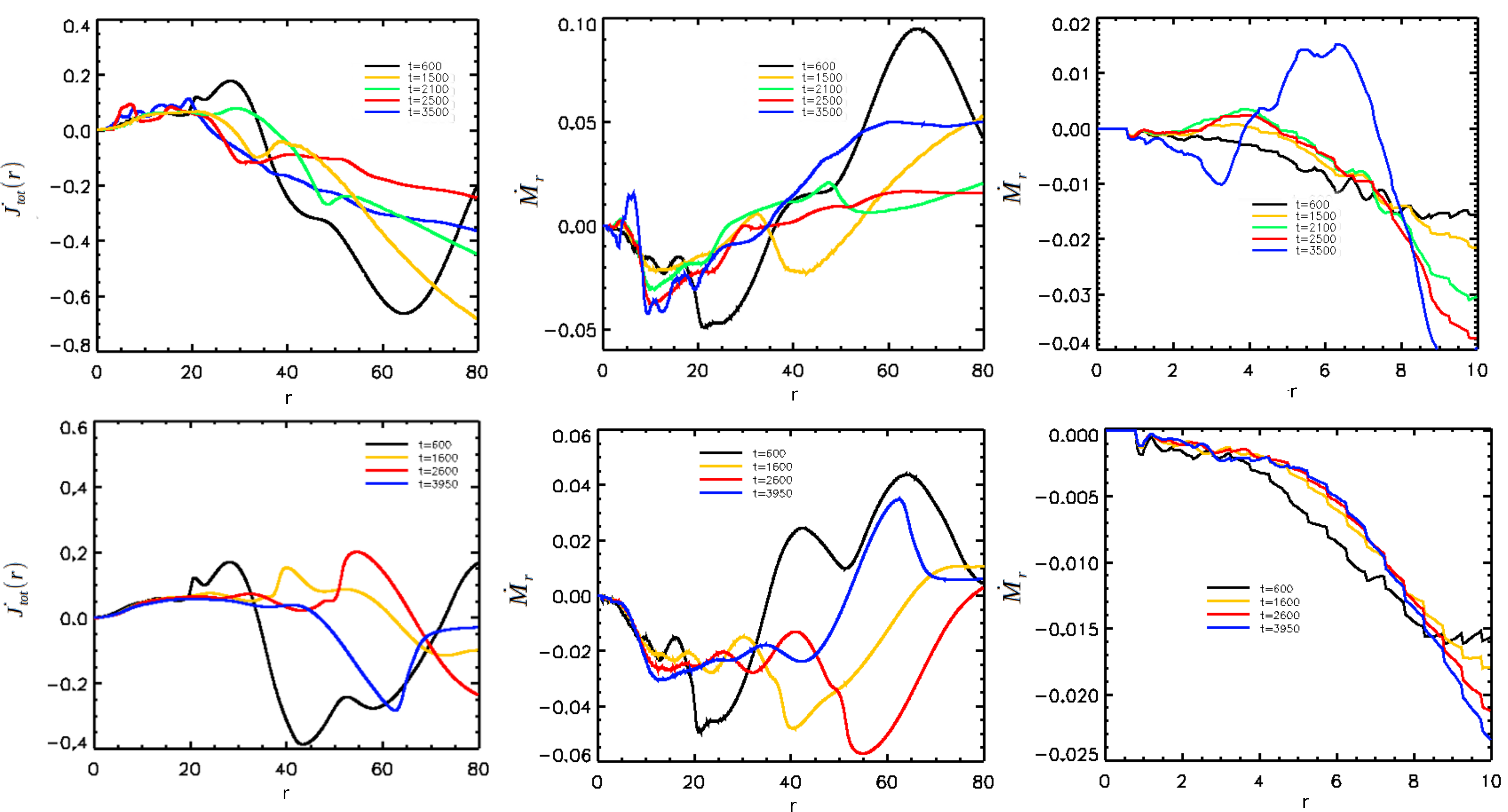}
\caption{Radial mass fluxes. Shown are the total angular momentum flux (first panel), the radial profile of $\dot M_r$ (second and third panel) for the binary star (top) and the single star (bottom) runs at different times.
} 
\label{mdot_rz_sin_bin}
\includegraphics[width=18cm]{\figurepath/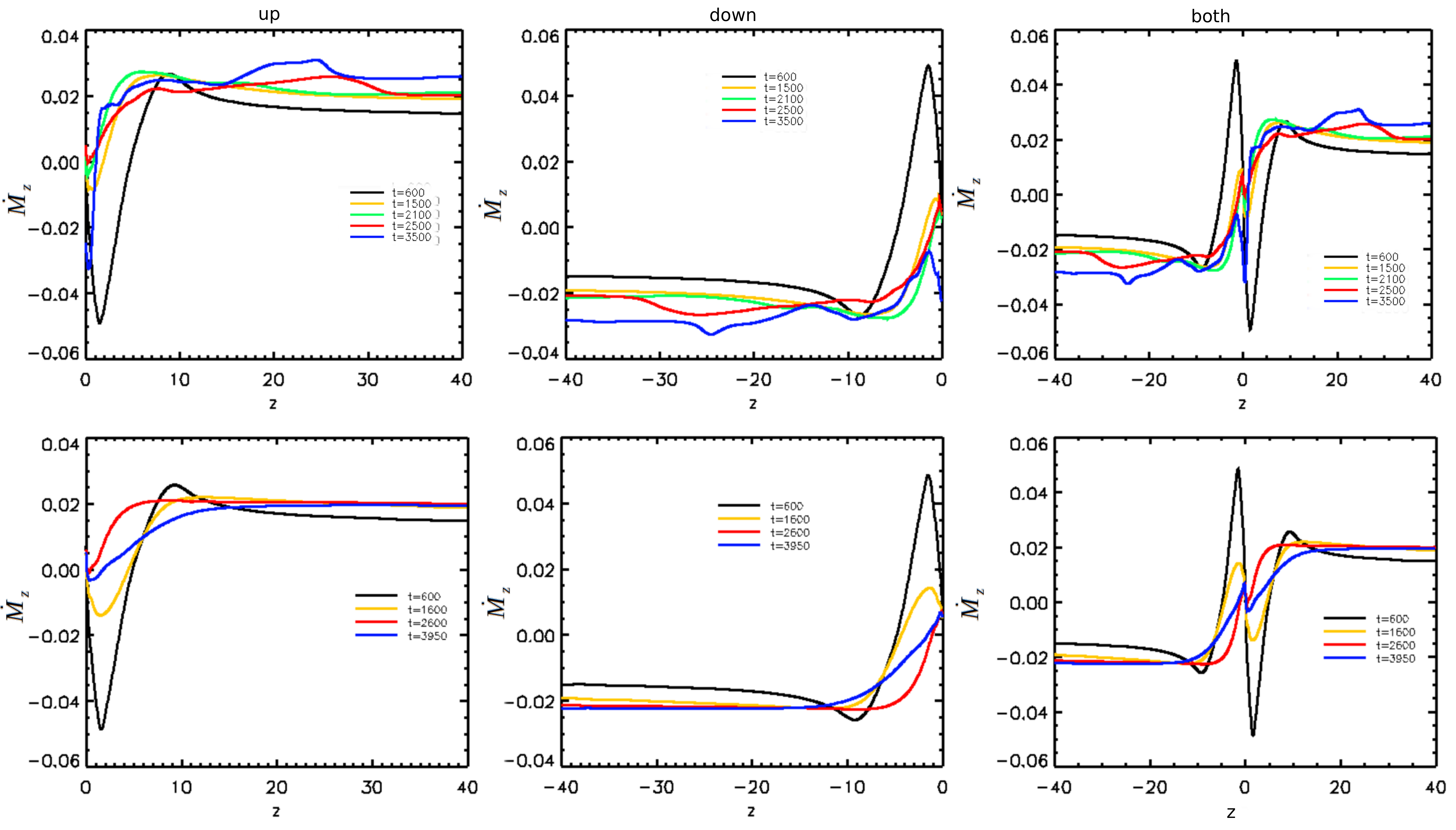}
\caption{Vertical mass fluxes. Shown are the vertical profile of $\dot M_z$ from up hemisphere (left panel), down hemisphere (middle panel) and both hemisphere (right panel), for the binary star (top) and the single star (bottom) runs at different times.
}
\label{fig:mdotz_both_hem}   
\end{figure*}

(i) We see that the most efficient driver in distributing the angular momentum in vertical direction 
is provided by the vertical motion of the magnetized outflow material, i.e.,$\dot{J}_2(z,t)$.
We also clearly see how the angular momentum flux of $\dot{J}_2(z,t)$ is correlated to the vertical
mass flux $\dot{M}_z$ (see Figure~\ref{mdot_rz_sin_bin}).

(ii) Similarly, we observe that the magnetic torque and the gravity torque are largest 
close to the disk and are decreasing for larger $z$.
This seems resulting from the fact that both the gravity and the magnetic field strength
are largest close to the disk.
In comparison to the disk area the magnetic torque gets smaller.

(iii) Considering the maps of the magnetic terms in Figure~\ref{tb_term_magnetic}, and also the radial and vertical profiles of 
the magnetic torque (Fig~\ref{jdotz_andt}), 
we find the term $l_{\rm B2}= (B_z / 4\pi\rho r) \partial_z(r B_{\phi})$ dominating.
The same is reflected in the integrated values for the radial ($\dot{J}_{br,2}$) and the vertical direction ($\dot{J}_{tbzi,2}$).
This is in nice agreement with the classical studies applying axisymmetry and only considering term $l_{\rm B2}$ \citep{2007prpl.conf..277P, 2019MNRAS.490.3112J}.
With our study, we confirm that term $l_{\rm B2}$ is dominant also in a non-axisymmetric treatment.
The contribution of the $\phi$-derivative of magnetic pressure term is minor.\\

We summarize this subsection by stating that among the additional terms considered in our model setup - compared to previous studies - the vertical motion contributes most for the vertical transport of angular momentum.
The 3D terms considering a $\phi$-variation of the physical variables, contribute relatively little to the global budget, probably since they average out when integrated over $\phi$. Nevertheless these terms vary by about 10-20\% along $\phi$. 

The largest impact results from the time-varying Roche potential, in particular for the areas inside the disk and close to the disk surface. 
We emphasize that, eventually, it is that variation that triggers all the non-axisymmetric effects we observe in the other physical terms.

\section{Overall fluxes of mass and angular momentum }
We finally investigate, how the efficiency of the global disk angular momentum transport 
is affected by the presence of the secondary star, thus by the action of its tidal torque.

We therefore integrate all local fluxes that have been discussed above in order to obtain
the global fluxes.
We compare the evolution in a binary system with the evolution of a single star system. 

A convenient parameter to evaluate the efficiency of the disk angular momentum transport is the mass flux transported inside the disk or the outflow.
We integrate the mass flux in radial direction as follows,
\begin{equation}
  \dot M(r,t) = \int_{-0.1 r}^{0.1r} \int_0^{2\pi} \rho r v_r \, d\phi \, dz.
  \label{radila mass flux}
\end{equation}
In Figure~\ref{mdot_rz_sin_bin} we display the mass fluxes in radial direction, $\dot{M}_r$.
The radial profile of the mass accretion rate $\dot{M}_r(r)$ is shown along the whole disk,
and also for the inner part of the disk (in higher resolution).

We see that for the setup of single star-disk-jet system the accretion process is smoothly 
established over time.
Accretion remains in action over the whole disk, also till the later evolutionary stages.
In particular, we find a negative accretion rate along the whole disk.
The accretion rate is converging to a value $\dot{M} \simeq 0.02-0.03$ (in code units), here
measured at a disk radius of $r \simeq 30$.

In contrast, we find that for the setup of a binary star-disk-jet system, disk accretion is severely 
affected by tidal effects - essentially visible as spiral arms in density and velocity as we have seen above.
Here, at late evolutionary stages we observe a change from accretion to an outward motion ({"}excretion{"}) for certain radii.
This is consistent with the distribution of radial velocities discussed earlier in Figure~\ref{fig:vr_binary} and ~\ref{fig:vr_single}.

The variation in the direction of mass flux does not happen at a fixed region and is
also evolving in time. 
For instance, at $t= 1500$ the transition from accretion to excretion is found at $r\simeq 30$.
This is close to the area of the predominant spiral structure, and also close to the
radius of the L1 Lagrange point (see Figure~\ref{fig:nc3_xy_rho_com}).
We conclude that different agents affect the variation in the disk accretion behavior. 
The most prominent ones arise from the evolution of the spiral arms, and from the orbital motion of the secondary (see position of the Lagrange point), altogether affecting the total angular momentum flux distribution in the disk.

It is certainly an intriguing questions whether the global disk accretion rates are affected by the 3D effects and the subsequent torques.
We may quantify this by considering the radial mass flux profiles of two different systems.
For late times, $t\simeq 3500$, and for radii of $r\simeq25$ (which is inside the accreting area of the disk in the binary setup),
we measure an average accretion rate of $-0.03$ (in code units) for the binary system, and of $-0.025$ (in code units) for the 
disk around the single star. 
This consequently implies that the accretion rates indeed may change due to the 3D effects discussed above, for the system parameters that we have investigated the radial mass fluxes increases by about $20\%$.

\begin{figure}
\centering
\includegraphics[width=1.0\columnwidth]{\figurepath/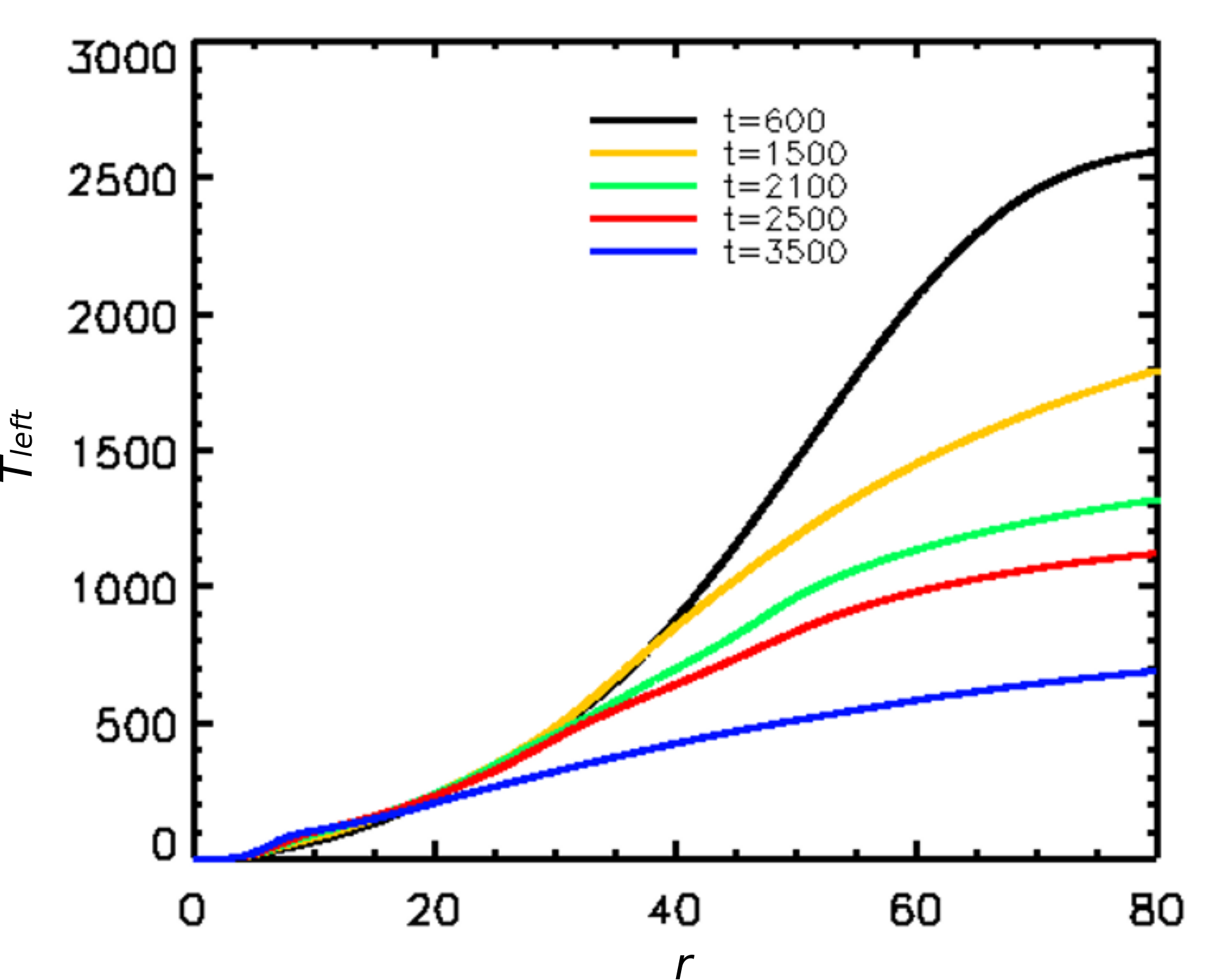} 
\caption{Test of the angular momentum budget.  Shown is the radial profile of the term at the left side of equation \ref{phi-torque2} denoted by ``$T_{\rm left}$''.}
\label{a.m budget_bin}
\end{figure}  

The mass fluxes in vertical direction we integrate as follows,
\begin{equation}
  \dot M(z,t) = \int_0^{80r_i} \int_0^{2\pi} \rho r v_z \, d\phi \, dr,
  \label{vertical mass flux}
\end{equation}
with the inner disk radius $r_{\rm i}$.
Figure~\ref{fig:mdotz_both_hem} shows the vertical profile of $\dot M_z$ from the upper hemisphere (left panel),
the lower hemisphere (middle panel) and both hemispheres together (right panel), 
for both the binary star simulation (top) and the single star simulation (bottom) at different times.

In contrast to the radial mass flux in the disk, we find a much smoother profile for the vertical
mass fluxes $\dot{M}(z,t)$. 
The profile for  vertical mass fluxes nicely 
demonstrate the transition from the accretion disk into the disk outflow. 
In particular, the profiles explicitly demonstrate the change of sign that takes place at the 
altitude of about one disk scale height, where radial mass accretion is diverted into a mass outflow 
as a disk wind.
The profiles also show how the disk outflow approaches a kind of steady state, saturating in a constant mass flux
in vertical direction, a result that is expected from steady-state MHD theory of disk 
winds\footnote{Note that steady-state MHD wind theory predicts conserved mass flux along magnetic flux 
surfaces, while in the Figures we plot the mass flux integrated over all flux surfaces.
}

Comparing the vertical mass fluxes for the binary and the single star simulation, we find very symmetric profiles for the 
run of the single star setup, which again highlights the quality of our simulation setup.
In contrast, the clear asymmetries seen in the profiles for the binary star simulation evidently demonstrate the influence of the tidal effects on the launching process caused by the secondary.

We may quantify this by considering the vertical mass flux profiles of the two setups at late times ($t\simeq 3500$).
Here, we measure average mass fluxes of 0.03 (in code units) for the binary star,
and 0.02 (in code units) for the single star. 
We conclude that disk winds in binary stars may carry 50\% more mass flux in comparison to a single star disk.

We also show the total angular momentum flux, $\dot{J}_{\rm tot}$ (see Figure~\ref{mdot_rz_sin_bin}, first panel).
It is helpful to stress again that in order to derive the radial profile of the total angular momentum fluxes at work, 
respectively the torques, we need to integrate each term (see Table~\ref{tbl:terms}, middle) 
in $r$-direction from $0$ to $r$.

For example, in order to obtain $\dot{J}_{tot}(r,t)$ at each radial position, we integrate
\begin{multline}
\dot{J}_{tot}(r,t) = \int_0^r \, [ Te_{\rm r,1}(r',t) + Te_{\rm r,2}(r',t) + Te_{\rm r,3}(r',t) \\ 
+ Te_{\rm r,4}(r',t) + Te_{\rm r,5}(r',t) + Te_{\rm r,6}(r',t) ] \,dr'
\end{multline}
Here, the control volume for integration covers the radii from $r=0$ to $r'=r$ and spans in vertical direction from 
$z=-0.1~r$ and $z=0.1~r$.
Thus, the term $\dot{J}_{\rm tot}(r,t)$ represents total angular momentum flux integrated along in $\phi$, in $z$ and in radial direction at each radius. 
In other words,  $\dot{J}_{tot}(r,t)$ shows the total angular momentum flux until that specific radius.

We find that the total angular momentum flux (respectively the net torques) at time $2500$ (in code units) is positive
for $r<50$, and changes sign for larger radii.
The radial profile of $\dot {J}_{tot}(r,t)$ allows to interpret the evolution of the disk angular momentum locally. 
The positive sign indicates that until this specific radius (r=50) the angular momentum of the disk material is removed from the inner regions
to larger radii.
Obviously, this supports accretion of matter.
However, for radii  $r>50$  we observe that removal of angular momentum does not take place. 
These areas in fact gain angular momentum, and, consequently, accretion turns into excretion.

More specifically, a parcel of mass that looses angular momentum (in any direction) will move inwards.
When it gains angular momentum, it will move outwards.
Now, if angular momentum continues to go outwards (for a certain range of radii), this part will constitute an {\em accretion} disk.
If angular momentum is transported inwards, this part of the disk may move in, however, the part of the disk within
this radius, has received angular momentum, and is supposed to move outwards.
Overall, this will not lead to a steady state situation.
We note that is why magnetic winds are so efficient for accretion as the vertical transport always removes angular momentum, thus leading to accretion for all radii.
In fact that is what we also observe on the long time scales.
The outer disk disappears and a smaller-size disk remains with a disk radius well within the radius of L1.

As a sanity check for the total the angular momentum budget of the binary star-disk-jet system, 
we compare both sides of the equation \ref{phi-torque2}.
We compute the term at the left side of the equation, now denoted by $T_{\rm left}$ and the total angular moment flux
$\dot {J}_{tot}(r,t)$ computed from the right side. 
Both terms are shown in Figure~\ref{mdot_rz_sin_bin} and Figure~\ref{a.m budget_bin}, respectively, and 
allow us to compare both sides of the angular momentum flux equation.
 
From these plots, we may consider, as an example, the values for $t_1 \simeq 1500$ (yellow) 
and $t_2 \simeq 2100$ (green) at $r=80$.
We first compute the left side of equation \ref{phi-torque2},
$\left( T_{\rm left}(t_2)  - T_{\rm left}(t_2)(t_1) \right) / \Delta t \simeq \dot J _{\rm tot}(t_1)$.
With $\Delta t = t_2 - t_1 = 640$ we calculate
$\left( T_{\rm left}(t_2)  - T_{\rm left}(t_1) \right)            \simeq 500$ and 

$\left( T_{\rm left}(t_2)  - T_{\rm left}(t_1) \right) / \Delta t \simeq 0.7$.

On the other hand, from the figures we also find $\dot J _{\rm tot}(t_1=1500) \simeq 0.7$ (yellow  line at $r=80$).
This nicely approves our angular momentum budget, as it shows the equivalence of the left and right side 
of equation \ref{phi-torque2} .

\section{Conclusions}
We have presented a detailed analysis of the angular momentum balance of the accretion-ejection system in a binary star.
For that we have re-visited our novel 3D MHD simulations that were published recently \citep{2018ApJ...861...11S}.
In particular, we have investigated how the existence of disk spiral arms influence the jet launching process and what kind of substructures emerge in the jets that evolve in our full 3D simulations.
We have further investigated to what extend the global properties, thus observables such as the disk accretion rate and jet mass flux are affected by the 3D effects, compared to a single-star accretion disk that launches an outflow. 

We have obtained the following results.

(i) As a general result for the evolution of the binary star-disk-jet system, 
we find that the initial disk size is decreasing and finally becomes confined to a size within the Roche lobe. As our model setup considers the full 3D evolution, we observe the growth of non-axisymmetric structures and spiral arms developing in the accretion disk.

(ii)
Considering the evolution of the disk spiral arms we recognize that the density wave and the pressure wave follow the same pattern speed.
It is clearly seen that with time the disk spiral arms become denser and more prominent and finally representing the main structural feature of the disk.
We find that the rotation of the spiral arms is synchronized with the orbital motion of the binary.
Furthermore, we see that the different sides of the disk experience different tidal forces --  resulting in a 
stronger and faster formation of spiral arm in that part of the disk that is closer to the L1.

(iii) The spiral arm pattern is also seen in the magnetic field structure. 
While the local differences of the corresponding magnetic torques cancel out when averaged over the full angle, 
they have an essential impact on the local launching conditions for the outflow.
In fact, they determine, together with density profile the particular 3D jet structure we observe (see below).

(iv)
Also for the velocity field of the system we find the same pattern in the disk that is involved in the formation of
the density spiral arm. 
In particular we find that the arm is not co-rotating with the material, but is synchronized with the orbital motion of the companion.

(v)
The velocity pattern observed in the binary simulation shows considerable differences compared to the 
one we observe in simulation for 3D jet launching from a single star.
The radial velocity pattern for the disk around a single star is similar to the typical accretion pattern
(thus a negative $u_r$) in almost axisymmetry.
In contrary, the accretion velocity for the binary star simulation looks quite unusual, exhibiting {"}excretion{"}
channels along certain angular directions.
These channels follow a spiral structure and are not aligned parallel the mid-plane.
Overall, we conclude that the radial velocity pattern seen in the binary disk is affected drastically by the tidal forces acting in the system.

(vi)
The azimuthal profile of the rotational velocity follows very closely the azimuthal profile of the disk density.
The peaks in the rotational velocity profile indicate the location of the spiral arm, while these peaks also 
indicate a very strong shear.
Thus, the enhanced {\em orbital} velocity that is present in the disk itself triggers further angular momentum exchange.

(vii)
As our central result, we find that the spiral structure of the disk is {\em launched into the jet outflow}.
Most prominently, these features are visible in the velocity pattern.
We may call these newly discovered jet structures {\em jet spiral walls}.
Essentially, the spiral features in the disk and in the jet follow the same kind of time evolution,
meaning that the jet spiral {}"walls{"} are establishing an almost stationary structure co-rotating with the disk.
We notice however that a small change in the position angle of the spiral structure along the outflow appears, 
resulting from the fact that the jet dynamical time scale is much faster than the disk dynamical time scale.

Thus, any structure that develops in the disk, is {"}immediately{"} propagated along the wind.
Nevertheless, on very large spatial scales we would expect the jet spiral arms to lag behind the disk 
spiral arms, assuming that the spiral structure and the jet survives that long.

(viii)
We investigated the global angular momentum budget in the binary star-disk-jet system and compared the respective 
impact of the particular physical terms.
In comparison to previous work, in our approach we have essentially considered the full magnetic torque 
and also the presence of an outflow, thus the angular momentum transport by vertical motion.

We find that among the extra terms we have considered that the vertical transport of angular momentum has a 
significant role in the total angular momentum budget also in a binary system.
The same holds for the magnetic torque, however, the contribution that arises from the $\phi$-derivative 
of the magnetic pressure (which is a truly 3D term) and the $B_{\phi}B_r$ stresses are small in the disk mid-plane. 
The gravity torque arising from the time evolution of the 3D Roche potential plays an essential role, as it
constitutes the fundamental cause for the all 3D effects appearing in our disk-jet system, 
including the spiral structure in the disk (seen in density but also the magnetic field) and in the outflow.\\

(ix)
From the radial profiles of the angular momentum fluxes we have concluded that the torques by gravity and the magnetic 
field remove the angular momentum from the disk and thus support the inward motion of the disk material (accretion).
However, the other contributions to the torques and the respective direction of transport
(radial or vertical angular momentum transport) do not show an unique behavior.
Depending on the radial position in the disk they may either remove or advect angular momentum of the disk.
The latter may lead to excretion of material at certain radii. \\

(x)
When comparing the radial and vertical mass fluxes and also the total angular momentum fluxes in the binary disk and in the disk around a
single star, we find that in the binary case, accretion is not supported throughout the whole disk, 
and the profiles for the angular momentum fluxes and radial mass fluxes are altered due to the tidal effects. 
Subsequently, we also detect the hemispherically asymmetric profiles of the vertical mass flux for the binary disk, also caused by tidal effects.
In comparison, for the single-star disk the evolution of the total angular momentum distribution supports the accretion process over the whole disk area and we find profiles for vertical mass fluxes that are perfectly symmetric for both hemisphere.
\\

In summary, from investigating in detail fully 3D MHD simulations of the launching process of jets from accretion disks that
are hosted by a binary star component, we disentangle a number of new dynamical features.
In particular, we see the disk spiral structure being ejected into the disk outflow and the jet, featuring {"}spiral walls{"} 
along the jet.
The different physical torques acting on the disk and the jet are all affected by the existence of a binary component,
thus changing in space and time along with the time-variation of the Roche potential.
The global observable parameters such as disk accretion rate and jet mass flux are substantially varied in comparison to a 3D single star 
launching situation.

\acknowledgements
We thank Andrea Mignone and the PLUTO team for the possibility to use their code.
We acknowledge really helpful comments by an unknown referee that have lead to a clearer presentation of our results.
Our simulations were performed on the ISAAC cluster of the Max Planck Institute for Astronomy
and the COBRA and DRACO clusters of the Max Planck Society.

\appendix 

\section{Comparison table for the different angular momentum terms involved}
For convenience, here we show the various terms that contribute to the angular momentum budget in the 
Equations~\ref{llintime}, \ref{phi-torque2}, and \ref{phi-torque_3vert}
and that are displayed in the various 
Figures~\ref{fig:eq16_disk}, \ref{fig:eq16_jet}, \ref{tb_term_magnetic}, 
\ref{jdotr_averged_binary}, \ref{jdotr_averged_single},
\ref{jdotz_andt}, \ref{jdotz_andtSingle},\ref{fig:mdotz_both_hem} and \ref{mdot_rz_sin_bin}.

\begin{table*} 
\caption{The various terms that contribute to the angular momentum budget in the different
equations and that are displayed in the various figures.
} 
\begin{center}
\begin{tabular}{cccc}
\hline
\hline
\noalign{\smallskip}
\multicolumn{4}{l}{Different terms contributing to Equation~\ref{llintime}. } \\
\noalign{\smallskip}
\hline
\noalign{\smallskip}
 ID   &  term                                      & physical meaning & \\
\noalign{\smallskip}
\hline
\noalign{\smallskip}
  $l_{\rm Ur}$ &  $  - u_r  \partial_r l$                    & radial transport   &  \\
  \noalign{\smallskip}
  $l_{\rm Uz}$ &  $ - u_z  \partial_z l$                    & vertical transport &  \\
  \noalign{\smallskip}
  $l_{\rm U\phi}$ &  $  - \frac{1}{r} u_\phi \partial_{\phi} l $  &  orbital transport &  \\
  \noalign{\smallskip}
  $l_{\rm G}$ &   $ - \partial_{\phi} \Phi_g$               &  gravity           &  \\
  \noalign{\smallskip}
  $l_{\rm P}$ &   $ - \frac{1}{\rho}  \partial_{\phi} P$    & gas pressure       &  \\
   \noalign{\smallskip}
  $l_{\rm B}$&   $   \frac{1}{\rho}   r F_{\rm B} $            & magnetic torque    &  \\
 \noalign{\smallskip}
  $l_{\rm B1}$&  $ \frac{1}{4\pi\rho r} B_{r} \partial_r (r B_{\phi}) $ & magnetic tension part 1 &  \\
  \noalign{\smallskip}
  $l_{\rm B2}$&  $ \frac{1}{4\pi\rho r} B_{z} \partial_z (r B_{\phi}) $ & magnetic tension part 2&  \\
  \noalign{\smallskip}
  $l_{\rm B3}$ & $ -\frac{1}{8\pi\rho r} \partial_\phi \left( B_r^2 + B_z^2 \right)$  & magnetic pressure gradient &  \\
\noalign{\smallskip}
\hline
\hline
\noalign{\smallskip}
\multicolumn{4}{l}{The different terms contributing to Equation~\ref{phi-torque2}. }  \\
\noalign{\smallskip}
\hline
\noalign{\smallskip}
 ID &   term & angular momentum fluxes (torques) & physical meaning  \\
\noalign{\smallskip}
\hline
\noalign{\smallskip}
 $\dot{J}_1(r,t)$ &  $ \int_0^r Te_{\rm r,1}(r',t) dr' $   & $ Te_{\rm r,1}=-\int\int \partial_r \left(\rho r u_r l\right) \,d\phi\,dz $  & radial transport\\
  \noalign{\smallskip}
 $\dot{J}_2(r,t)$ &  $\int_0^r Te_{\rm r,2}(r',t) dr'$ &$ Te_{\rm r,2}=-\int\int \partial_z \left(\rho r u_z l\right) \,d\phi\,dz $   & vertical transport \\
  \noalign{\smallskip}
 $ \dot{J}_3(r,t)$ & $ \int_0^r Te_{\rm r,3} (r',t) dr'$   & $Te_{\rm r,3} =-\int\int \partial_\phi \left(\rho u_\phi l\right) \,d\phi\,dz$ & orbital transport\\
  \noalign{\smallskip}
 $\tau_{\rm G}(r,t)$ & $\int_0^r Te_{\rm r,4}(r',t) dr' $  & $Te_{\rm r,4} =-\int\int r\rho \partial_\phi \Phi \,d\phi\,dz $ & gravity torque\\
  \noalign{\smallskip}
 $\tau_{\rm P}(r,t)$ &  $\int_0^r Te_{\rm r,5} (r',t) dr' $  &  $Te_{\rm r,5}=-\int\int r \partial_\phi P \,d\phi\,dz $& pressure torque \\
  \noalign{\smallskip}
 $\tau_{\rm B}(r,t)$ &$\int_0^r  Te_{\rm r,6}(r',t) dr' $ &   $ Te_{\rm r,6}=\int\int r^2F_{\rm B} \,d\phi\,dz $  & magnetic torque \\
\noalign{\smallskip}
$\dot J_{\rm br,1}(r,t)$  &$\int_0^r Te_{\rm B,r,1} (r',t) dr' $  & $Te_{\rm B,r,1}= \int\int \frac{1}{4\pi\rho r} B_{r} \partial_r \left(r B_{\phi}\right) \,d\phi\,dz $  &  magnetic tension part 1 \\
\noalign{\smallskip}
 $ \dot J_{\rm br,2}(r,t)$  &$\int_0^r Te_{\rm B,r,2} (r',t) dr' $   &  $Te_{\rm B,r,2}= \int \int  \frac{1}{4\pi\rho r} B_{z} \partial_z \left(r B_{\phi} \right) \,d\phi\,dz $ & magnetic tension part 2 \\
\noalign{\smallskip}
$\dot J_{\rm br,3}(r,t)$  & $\int_0^r Te_{\rm B,r,3} (r',t) dr' $  & $Te_{\rm B,r,3} =\int \int -\frac{1}{8\pi\rho r} \partial_\phi \left(B_r^2 + B_z^2\right) \,d\phi\,dz $  & magnetic pressure gradient \\
\noalign{\smallskip}
\hline
\hline
\noalign{\smallskip}
\multicolumn{4}{l}{Different terms contributing to Equation~\ref{phi-torque_3vert}.}\\
\noalign{\smallskip}
\hline
\noalign{\smallskip}
 ID & term & physical meaning & \\
  \noalign{\smallskip}
 \hline
$\dot J_1(z,t)$ & $ -\int\int \partial_r \left(\rho r u_r l\right) \,d\phi\,dr  $  &  radial transport &      \\
  \noalign{\smallskip}
$\dot J_2(z,t) $ & $  -\int\int \partial_z \left(\rho r u_z l\right)\, d\,\phi\,dr$ & vertical transport  &\\
  \noalign{\smallskip}
$  \dot J_3(r,t)$ & $ -\int\int \partial_\phi \left(\rho u_\phi l\right) \,d\phi\,dr $ &  orbital transport   &       \\
  \noalign{\smallskip}
$\tau_{\rm G}(z,t)$ & $ -\int\int r\rho \partial_\phi \Phi \,d\phi\,dr$  & gravity torque   &         \\
  \noalign{\smallskip}
$ \tau_{\rm P}(z,t)$ &  $ -\int\int r \partial_\phi P \,d\phi\,dr $    & pressure torque   &   \\
  \noalign{\smallskip}
$\tau_{\rm B}(z,t) $  & $ \int\int r^2 F_{\rm B} \,d\phi\,dr  $   & magnetic torque  & \\
\noalign{\smallskip}
\hline
\noalign{\smallskip}
\multicolumn{4}{l}{We note that the first integration in vertical direction gives us the angular momentum flux along $z$, thus  }\\
\multicolumn{4}{l}{we do not need to involve a further integration.}\\
 \end{tabular}
 \end{center}
\label{tbl:terms}
\end{table*}

\section{Comparison to a single star accretion disk-jet structure}
It is essential to test to code setup and also the plotting routines by cases for which we know the outcomes.
We have therefore run a 3D simulations of MHD jet launching for a single star.
Here we show a few exemplary results for comparison to the results of the binary star setup.

\begin{figure*}
\centering
\includegraphics[width=18cm]{\figurepath/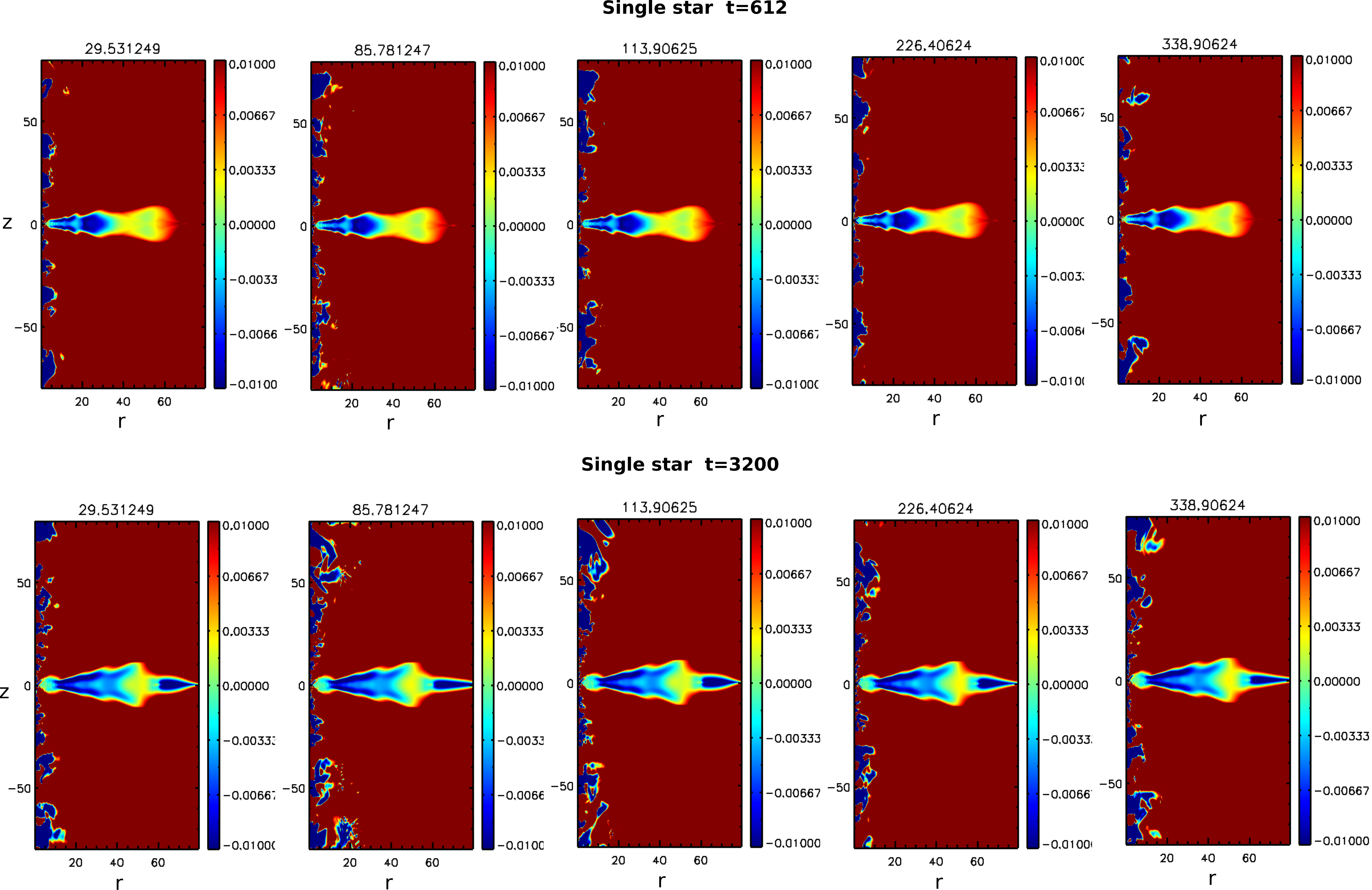}
\caption{Accretion-ejection velocity field. Shown are the snapshots of the radial velocity $u_r$ for a run with the single tar.
The angle $\Phi = 0$ is measured from the $x$-axis. 
Different $r$-$z$ planes correspond to different angles $\phi$.
Colors are enhanced to demonstrate inflow (disk) and outflow (jet)
}
\label{fig:vr_single}
\end{figure*}

In Figure~\ref{fig:vr_single} we shows the radial velocity distribution for the single star setup,
in particular enhanced for the accretion velocity.
This has to be compared to Figure~\ref{fig:vr_binary} in the main text for the binary star simulation. 

\begin{figure*}
 \centering
\includegraphics[width=18cm]{\figurepath/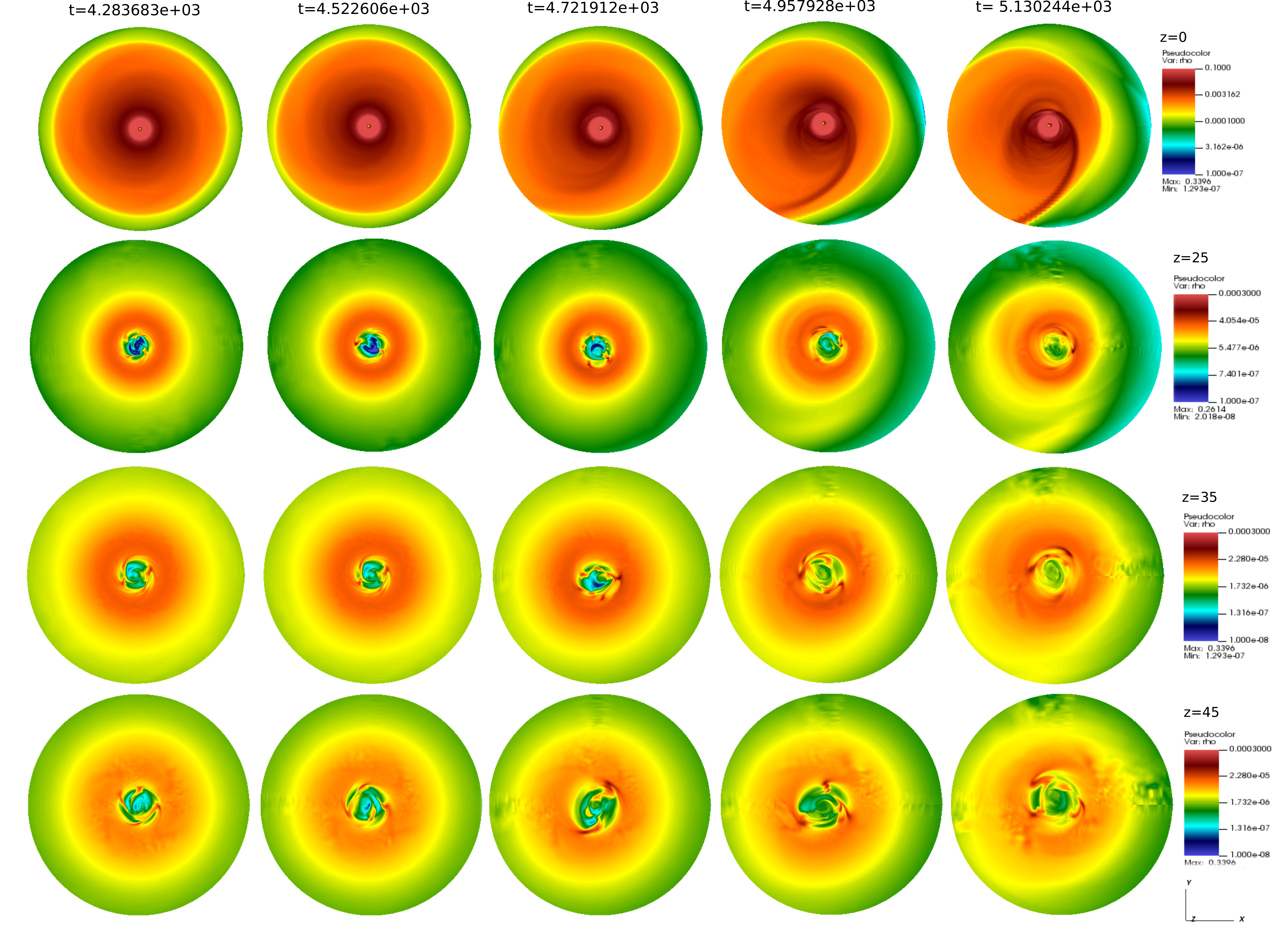} 
\caption{Origin of the jet spiral arms.  
This figure shows snapshots of mass density at different cross sections (at different z) and different times. 
In this simulation, we start from a single star setup that is run till $t=4228$,  
and then switch on the time-dependent Roche potential. 
Essentially, we see the formation of spiral arm starts in the disk earlier than for the jet area. 
Note, the small scale instabilities along the inner jet, that must be distinguished from the large
scale evolution of a spiral arm structure seen later.
}
\label{sin_to_bin_spiralwall}
\end{figure*}
\begin{figure*}
 \centering
\includegraphics[width=18cm]{\figurepath/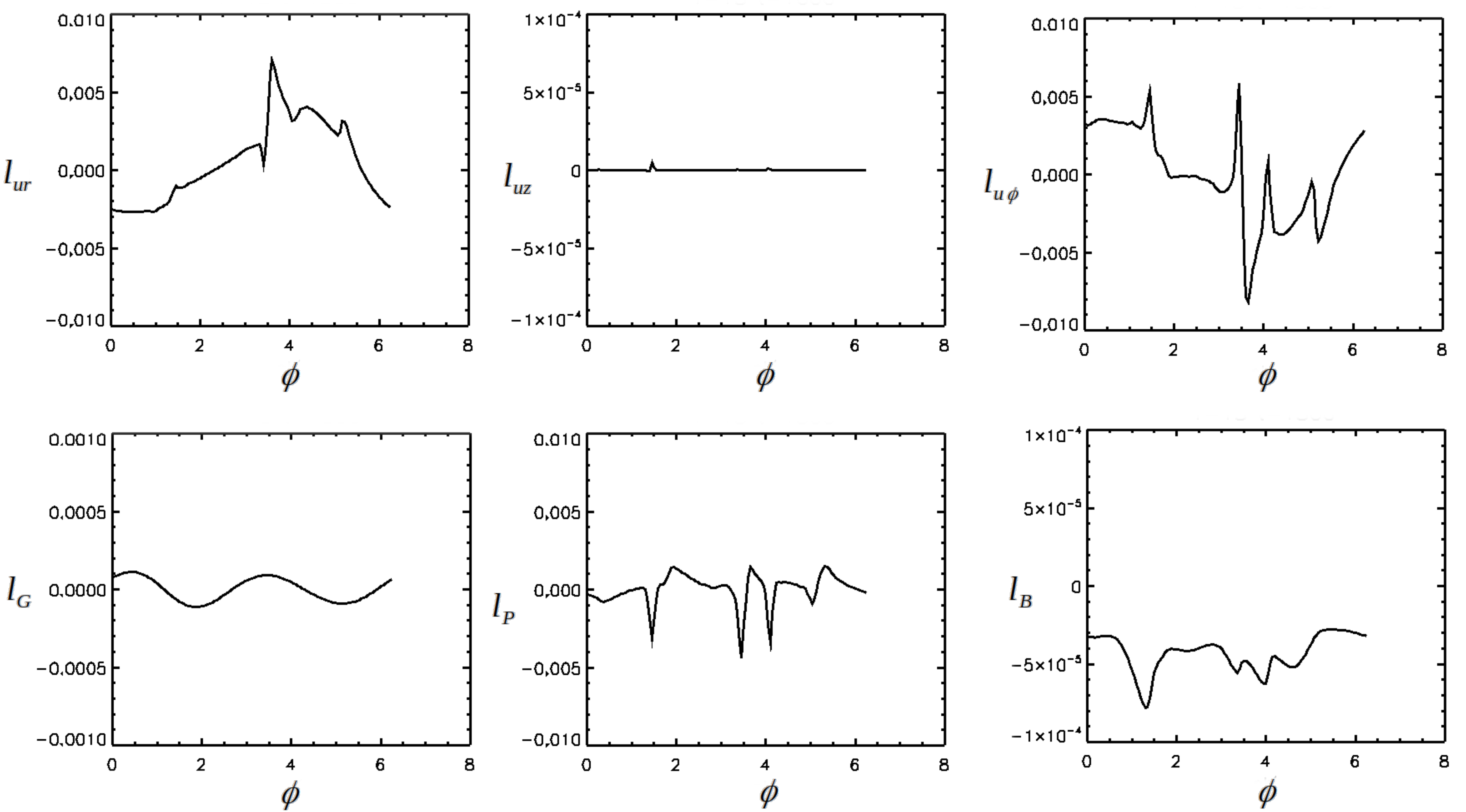}
\caption{Shown are the profiles of the specific angular momentum contributing in Equation~\ref{llintime} along $phi$ direction
for a simulation run of a disk in a binary system, at time 1500.}
\label{te_eq16_randphi_bin}
\includegraphics[width=18cm]{\figurepath/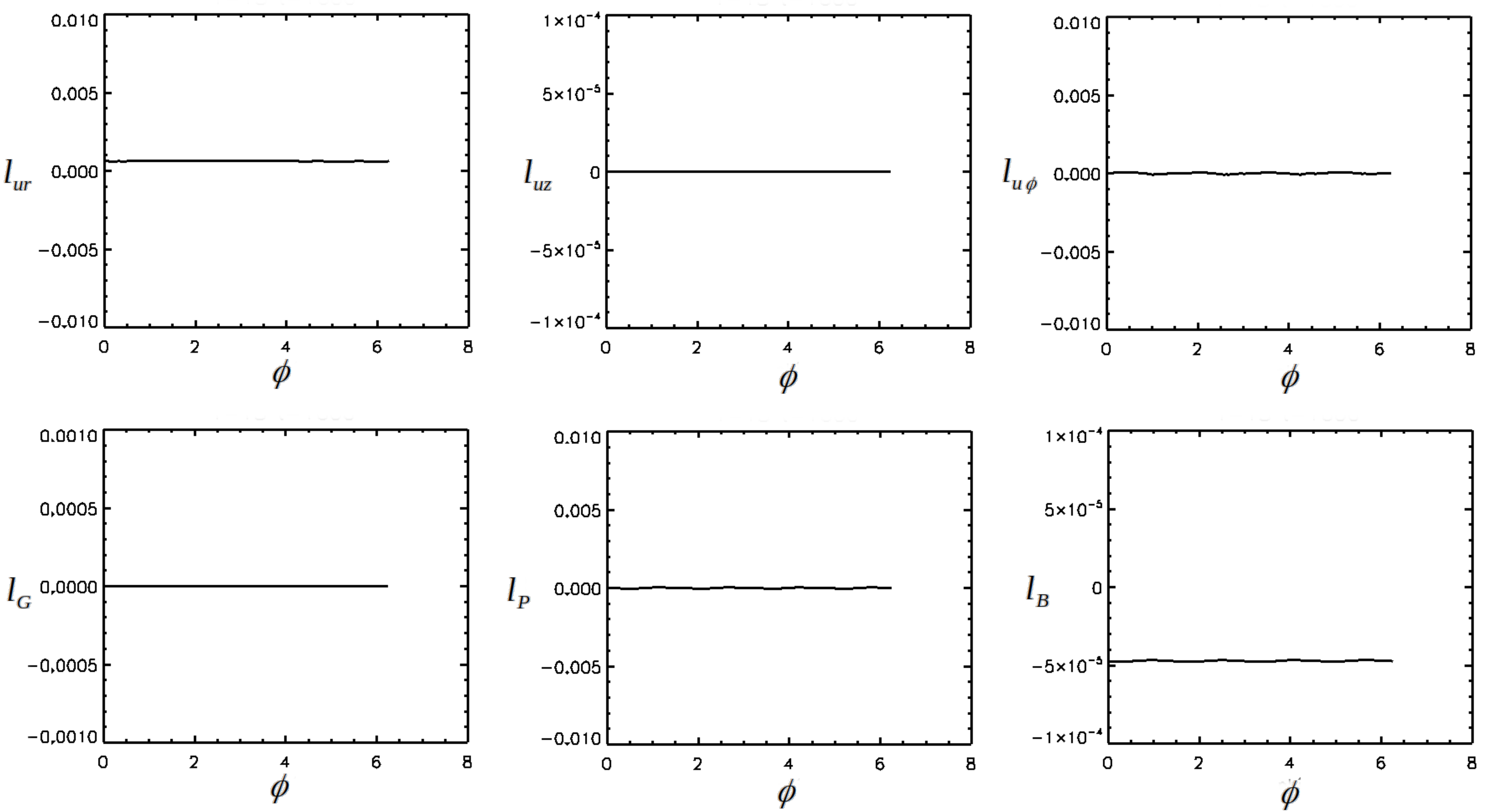}
\caption{Shown are the profiles of the specific angular momentum contributing in Equation~\ref{llintime} along $phi$ direction
for a simulation run of a disk around a single star, at time 1600.}
\label{te_eq16_randphi_sngl}
\end{figure*}

With Figures~\ref{te_eq16_randphi_bin} and \ref{te_eq16_randphi_sngl} we compare the strength of the terms in 
Equation~\ref{llintime} for the case of the binary star simulation and the single star simulation. For a discussion see the main text.

\section{Units and normalization}
\label{app:units-normalization}
We normalize all variables, in particular $r, \rho, u, B$ to their fiducial values at the inner disk 
radius $R_{\rm i}$.
For the astrophysical scaling, we have adopted the following number values for 
a young stellar object (YSO) of $M = 1\,M_{\odot}$ and 
an active galactic nucleus (AGN) of $M = 10^8\,M_{\odot}$.
Any change in one of the system parameters accordingly changes the scaling of our simulation results
which can thus be applied for a variety of jet sources.
The inner disk radius is usually assumed to be a few radii of the central object,
\begin{eqnarray}
 R_{\rm i} & = & 0.028\,{\rm AU} \left( \frac{r_{\rm i}}{3 R_{\rm YSO}} \right)
                             \left( \frac{R_{\rm YSO}}{2 R_{\odot}} \right) \quad\quad {\rm (\rm YSO)} \nonumber \\
           & = & 10^{-4}\,{\rm pc} 
                             \left( \frac{r_{\rm i}}{10 R_{\rm S}} \right)
                             \left( \frac{M}{10^8 M_{\odot}} \right) \quad \quad (\rm AGN),
\end{eqnarray}
where $R_{\rm S} = 2GM/c^2$ is the Schwarzschild radius of the central black hole. 
For relativistic sources, the inner disk radius is usually assumed to be located at the marginally 
stable orbit at $3\,R_{\rm S}$ (depending on the black hole rotation)
Since we apply the non-relativistic version of the PLUTO code, we cannot treat any relativistic 
effects. 
We may therefore consider a scaling $R_{\rm i} \simeq 10 R_{\rm S}$.
For simplicity and a convenient comparison with the other literature we apply $R_{\rm i} = 0.1\,$AU
for most comparisons concerning stellar sources.
The orbital velocity at the inner disk radius is
\begin{eqnarray}
 u_{\rm k,i} & = & 180\,{\rm km\,s^{-1}}\,
                 \left( \frac{M}{M_\odot} \right)^{1/2}
                 \left( \frac{r_{\rm i}}{3 R_{\rm YSO}} \right)^{-1/2} 
                 \left( \frac{R_{\rm YSO}}{2 R_{\odot}}  \right)^{-1/2}      \quad \quad {\rm (YSO)} \nonumber \\
             & = & 6.7\times 10^{4}\,{\rm km\,s^{-1}}
                 \left( \frac{r_{\rm i}}{10 R_{s}} \right)^{-1/2}
                 \left( \frac{M}{10^8 M_{\odot}}   \right)^{-1/2}    \quad \quad (\rm AGN).
\end{eqnarray}
For a $R_{\rm i} = 0.1\,$AU distance from the central star the orbital speed is 
$v_{\rm k,i} = 94\,\rm km\,s^{-1}$.
The mass accretion rate is a parameter which is in principle accessible by observation.
As a consequence, subject to the disk model applied, the observed disk luminosity can be related 
to an accretion rate. 
For a YSO, the accretion rate is typically of the order of 
$\dot{M}_{\rm acc} \simeq 10^{-7} {\rm M_{\odot} yr^{-1}}$,
providing the scaling of the density $\rho_{\rm i}$ applying 
$\dot M_{\rm i} = r_{\rm i}^2 \rho_{\rm i} u_{\rm K,i}$.
For a length scale $R_{\rm i} = 0.1\,$AU and a velocity scale  $u_{\rm k,i} = 94\,\rm km\,s^{-1}$, 
we obtain
\begin{eqnarray}
 \dot{M}_{\rm i} & = & 10^{-5} {\rm M_{\odot} yr^{-1}} 
                      \left( \frac{\rho_{\rm i}}{10^{-10} \rm g\,cm^{-3}} \right)
                      \left( \frac{M}{M_{\odot}} \right)^{1/2} 
                      \left( \frac{r_{\rm i}}{0.1\,\rm AU} \right)^{3/2}   \quad\quad {\rm (YSO)} \nonumber \\
            & = & 10\,{\rm M_{\odot} yr^{-1}}
                       \left(\frac{\rho_0}{10^{-12} \rm g cm^{-3}}\right)
                        (\frac{M}{10^8 M_{\odot}})^{1/2}           \quad\quad (AGN)
 \end{eqnarray}
Considering the normalization of the velocity by Keplerian velocity and the density by the $\rho_i$ (both at the inner radius of the disk),
the gas pressure is normalized according to the sound speed, i.e., $p_{i} = \epsilon^2 \rho_{i} u_{k,i}^2$ and here $\epsilon = c_s/u_{k} $ is the initial aspect ratio of the disk.

The scaling of the magnetic field is then obtained by considering the plasma-$\beta$ and the field 
strength at the equator at the inner radius $B_{\rm i}= \sqrt{8 \pi P_{\rm i}/\beta_{\rm i} } $,  
\begin{eqnarray}
 B_{\rm i} & = & 14.9 \left(\frac{\beta_{\rm i}}{10}\right)^{-1/2}
                  \left(\frac{\epsilon}{0.1}\right) \left(\frac{\rho_0}{10^{-10} \rm g cm^{-3}}\right)^{1/2}
                  \left(\frac{M}{M_\odot}\right)^{1/2} \left(\frac{r_{\rm i}}{\rm 0.1 AU}\right)^{-5/4}
                  \quad\quad {\rm G \quad (YSO) }\\
               & = & 1.06\times10^3 \left(\frac{\beta_{\rm i}}{10}\right)^{-1/2}\left(\frac{\epsilon}{0.1}\right) 
                  \left(\frac{\rho_0}{10^{-12} \rm g cm^{-3}}\right)^{1/2}
                  \left(\frac{r_{\rm i}}{\rm 10 R_{\rm s}}\right)^{-5/4} 
                  \quad\quad {\rm G \quad (AGN) }
\end{eqnarray}

\bibliographystyle{apj}
\bibliography{bibpaper2021}

\end{document}